\documentclass[iop]{emulateapj}

\newcommand{\asiaa}{Institute of Astronomy and Astrophysics, Academia Sinica, Taipei, 10617, Taiwan}
\newcommand{\tj}{T_\mathrm{J}}
\newcommand{\anep}{a_\mathrm{Nep}}
\newcommand{\rhn}{R_\mathrm{HN}}
\newcommand{\asj}{a_\mathrm{J}}
\newcommand{\njfc}{N_\mathrm{JFC}}
\newcommand{\ntot}{N_\mathrm{tot}}
\newcommand{\tsim}{t_\mathrm{sim}}
\newcommand{\nsrc}{N_\mathrm{src}}
\newcommand{\rr}{r_\mathrm{R17}}
\newcommand{\rsrc}{r_\mathrm{src}}

\usepackage{graphicx}
\usepackage[caption=false]{subfig}
\usepackage{amsmath}
\usepackage{hyperref}
\usepackage{xcolor}

\slugcomment{Accepted in AJ}

\shorttitle{DPs Contribution to the Origin of JFCs}
\shortauthors{Mu\~noz-Guti\'errez et al.}

\begin{document}


\title{The Contribution of Dwarf Planets to the Origin of Jupiter Family Comets.}



\author{M. A. Mu\~noz-Guti\'errez}
\affil{\asiaa}
\email{mmunoz@asiaa.sinica.edu.tw}
\author{A. Peimbert, B. Pichardo}
\affil{Instituto de Astronom\'ia, Universidad Nacional Aut\'onoma de 
M\'exico, Apdo. postal 70-264, Ciudad Universitaria, M\'exico}
\author{M. J. Lehner}
\affil{\asiaa}
\affil{Department of Physics and Astronomy, University of Pennsylvania, 209 S. 33rd St. Philadelphia, PA 19104}
\affil{Harvard-Smithsonian Center for Astrophysics, 60 Garden St., Cambridge, MA 02138}
\author{S-Y. Wang}
\affil{\asiaa}



\begin{abstract}

We explore the long-term evolution of a bias-free orbital representation of the cometary nuclei (with diameters above 2 km) of the Kuiper belt, using the so-called L7 synthetic model from CFEPS, which consists of three dynamical sub-populations: the Classical, the Resonant, and the Scattering. The dynamical evolution of belt particles is studied under the gravitational influence of the Sun and the four giant planets, as well as of the 34 largest known trans-Neptunian objects (TNOs with $H_V<4$). Here we indistinctly call Dwarf Planets (DPs) to the full sample of 34 large TNOs. Over a 1 Gyr time-scale, we analyze the secular influence of the DPs over Kuiper belt disk particles and their contribution to the injection rate of new visible Jupiter Family Comets (JFCs). We find that DPs globally increase the number of JFCs by 12.6\%, when compared with the comets produced by the giant planets alone. When considering each population separately, we find that the increment produced by DPs is 17\%, 12\%, and 3\% for the Classical, Resonant, and Scattering populations, respectively. Given the rate of escapes from the Kuiper belt, we find upper limits to the number of objects in each population required to maintain the JFCs in steady state; the results are $55.9\times10^6$, $78.5\times10^6$, and $274.3\times10^6$ for the Scattering, Resonant, and Classical populations, respectively. Finally, we find that the Plutinos are the most important source of comets which were originally in a resonant configuration, where the presence of Pluto alone enhances by 10\% the number of JFCs. 

\end{abstract}


\section{Introduction}

Among all the comets observable to date, two broad categories have been established based on their dynamical properties and orbital distributions. The long period comets (LPCs), or nearly-isotropic comets, on one side, are comets with orbital periods greater than 200 yr which are homogeneously distributed around the Sun, i.e. their inclinations can cover all possible values from 0 to 180$^\circ$. The distribution of LPCs strongly indicates the existence of a spherically distributed big reservoir located at large distances around the Sun, i.e. the Oort Cloud \citep{Oort50,Weissman90,Wiegert99,Dones04,Kaib09}.

On the other hand, the so-called ``ecliptic comets'' are a distinctive type of comets with a narrower inclination distribution, concentrated at low values; thus they remain close to the ecliptic plane at all times. The orbital periods of ecliptic comets are typically smaller than 200 yr, in fact, the vast majority of them have periods below 20 yr.

A dynamical classification of comets, based on the value of their Tisserand parameter with respect to Jupiter, $\tj$, has been suggested to quantitatively separate the comet families \citep{Levison96}. In this scheme, the ecliptic comets are easily distinguished from the LPCs: the ecliptic comets have values of $\tj>2$ while the LPCs have values of $\tj<2$ \citep[see also][]{Carusi87,Duncan04,Tancredi14}. Further subdivisions can be established among the ecliptic comets based on their values of $\tj$; objects that at present are dynamically dominated by Jupiter (i.e. Jupiter Family Comets, JFCs) have $2<\tj<3$. As of February 2019, JFCs constitute approximately the 80\% of all the short period comets (SPCs, those with $P<200$ yr), according to the JPL Small-Body Database Search Engine\footnote{\url{https://ssd.jpl.nasa.gov/sbdb_query.cgi} Note however, that such figures most likely are not representative of the intrinsic comet population in the solar system, as a strong discovery bias is present for objects that get closer to the Sun during perihelia.}. 

The origin of the comets we observe, in the inner part of the Solar System, has been a longstanding problem that has received the attention of astronomers since early times. JFCs in particular, with their narrow distribution of inclinations, represent a distinct family of objects with a specific reservoir. Early attempts to reconcile the idea that JFCs come from the Oort cloud showed that, although possible, such a dynamical path is in extreme inefficient at producing low inclination comets in short-period orbits, thus showing that such scenario is unlikely to be viable \citep{Everhart72,Joss73,Fernandez94}. In a seminal paper \citet{Fernandez80} \citep[working on the ideas of][]{Edgeworth49,Kuiper51,Kuiper74} was the first to numerically examine the possibility of the existence of a comet belt beyond Neptune in connection with the origin of the SPCs. Such a reservoir would be a much more suitable source for the production of SPCs in low inclination orbits than the Oort cloud. Later works soon confirmed, by means of full numerical simulations, the feasibility of the trans-Neptunian comet belt as the source of JFCs \citep{Duncan88,Torbett89,Quinn90,Ip91}, thus giving origin to the ---then still theoretical---  Kuiper belt \citep{Quinn90}.

After the first member of the Kuiper belt (other than Pluto) was discovered \citep{Jewitt93}, hundreds of new objects have been observed in the trans-Neptunian region thanks to several dedicated surveys \citep[see for instance][]{Jewitt98,Trujillo01,Millis02,Petit11,Bannister18}. The new observational data revealed that the orbital structure of the Kuiper belt is much more complex than originally expected; this realisation drove a new understanding on the early evolution of the Solar System, and also drew the picture of the origin and dynamical evolution of Solar System's cometary reservoirs \citep[see][for a recent review]{Dones15}.

Our current understanding suggests that the early migration of the giant planets, driven by interactions with leftover planetesimals, was responsible for generating the orbital structure currently observed in the Kuiper belt. Such structure is characterized by the overpopulation of Neptune's mean motion resonances (MMRs), the dynamical excitation of the Classical Kuiper belt (which is composed of a hot and a cold population), and the existence of a Scattered disk. In addition to these, it is argued that an important part of the leftover plenetesimals, in the original protoplanetary disk, ended up forming the Oort cloud, after they were scattered off by the migrating giant planets.

After the setting of the giant planets in their current orbits, the secular evolution of the cometary reservoirs, under the long-lasting planetary configuration, proceeded for approximately 4 Gyr. Under the current conditions, it has been shown that perturbations produced by the giant planets alone are able to supply the number of new objects required to maintain the population of JFCs in steady state, provided that a large enough population of cometary nuclei are present in the trans-Neptunian reservoir regions. Different authors have considered different components as the sole reservoir of the JFCs: either the Classical Kuiper belt \citep[e.g.][]{Duncan95,Levison97,Nesvorny17}, the populations in MMRs, mainly the Plutinos \citep[e.g.][]{Morbidelli97,Ip97} or the Scattered disk \citep[e.g.][]{Duncan97,Emelyanenko04,Volk08,DiSisto09}.

Despite those important efforts to conciliate the above picture with the number of JFCs currently present in the inner solar system, some discrepancies still remain. In a recent work, \citet{Nesvorny17}, considering an end-to-end numerical scenario since the formation of the cometary reservoirs and up to 4 Gyr of dynamical evolution, found that the predicted number of ecliptic comets still falls short by a factor of at least two when compared with observations, unless a size-dependent physical evolutionary model is taken into account.

In this work we consider an additional source of perturbations known to exist in the trans-Neptunian region, one which has been typically underestimated: the dwarf planets (DPs). Some previous works had focused on the effect that Pluto exerts on the long-term evolution of the Plutinos \citep{Nesvorny00,Tiscareno09}, however, we have previously shown that a large group of DPs is able to destabilize ---in a significant measure--- the orbits of cometary material in debris disks \citep{Munoz15,Munoz17,Munoz18}.

Here we use an unbiased orbital representation of the Kuiper belt ---the L7 model, released by the CFEPS team--- to numerically study the effect that the 34 largest TNOs (observed to date) exert over the cometary population of the belt. We found that the large objects have a significant effect on long-term time scales, leading to a modest enhancement of the injection rate of new comets into the inner solar system. The reservoir requirements implied by our injection rate remain in broad agreement with recent reservoir estimations done from observations of the trans-Neptunian regions, as well as with those estimated from the analysis of cratering records on Pluto and 2014~MU$_{69}$  \citep{Greenstreet15,Greenstreet19,Singer19}.

Tno los pued..
his paper is organized as follows: in Section \ref{sec:sims} we describe the numerical simulations of the Solar System performed, as well as the population of DPs and cometary nuclei that we used for the simulations. Section \ref{sec:res} is devoted to present and discuss the results for the different populations of the L7 model, with and without DPs. In Section \ref{sec:rates} we present the in-falling rates of JFCs as well as the number of objects in the trans-Neptunian region required to maintain the comet population in steady state. Finally, in Section \ref{sec:conc} we present our main conclusions.

\section{Simulations of the Solar System}
\label{sec:sims}

\subsection{The sample of the largest TNOs in the Solar System.}
\label{ss:tnos}

\begin{deluxetable}{lcr@{}lc}
\tabletypesize{\scriptsize}
\tablecaption{The sample of the 34 largest TNOs in the Solar System used in this work.\label{tbl:allDPs}}
\tablewidth{0pt}
\tablehead{
\colhead{Name} & \colhead{$R$ (km)} & \multicolumn{2}{c}{$\rho$ (gr cm$^{-3}$)} & \colhead{$M$ ($\times 10^{-3}$ M$_\oplus$)}
}
\startdata
Eris & 1200.0 & \null\hspace{10pt}2.&3 & 2.7956 \\
Pluto & 1188.3 & 1.&854 & 2.4467 \\
Makemake & 717.0 & 2.&14 & 0.5531 \\
Haumea & 806.8 & 1.&821 & 0.6706 \\
2007 OR$_{10}$ & 767.2 & 1.&9273 & 0.6110 \\
Orcus & 450.0 & 1.&676 & 0.1073 \\
Quaoar & 551.7 & 1.&99 & 0.2343 \\
Varda & 371.5 & 1.&24 & 0.0446 \\
2007 UK$_{126}$ & 324.0 & 1.&74 & 0.0415 \\
2002 TX$_{300}$ & 143.0 & 0.&933 & 0.0018 \\
2002 UX$_{25}$ & 332.0 & 0.&82 & 0.0209 \\
Varuna & 334.0 & 0.&992 & 0.0259 \\
2003 AZ$_{84}$ & 385.5 & 0.&87 & 0.0349 \\
\\
Sedna & 497.5 & 1.&4532 & 0.1254  \\
2002 AW$_{197}$ & 384.0 & 1.&5277 & 0.0606 \\
2014 UZ$_{224}$ & 317.5 & 1.&1865 & 0.0266 \\
2005 UQ$_{513}$ & 249.0 & 1.&1206 & 0.0121 \\
Ixion & 308.5 & 1.&2811 & 0.0263 \\
2002 MS$_{4}$ & 467.0 & 1.&9176 & 0.1369 \\
2005 QU$_{182}$ & 208.0 & 1.&3957 & 0.0088 \\
2015 RR$_{245}$ & 335.0 & 1.&2577 & 0.0331 \\
2005 RN$_{43}$ & 339.5 & 0.&9326 & 0.0255 \\
2002 TC$_{302}$ & 292.0 & 1.&3725 & 0.0239 \\
2015 KH$_{162}$ & 400.0 & 1.&3236 & 0.0594 \\
2010 EK$_{139}$ & 235.0 & 0.&6002 & 0.0054 \\
2004 GV$_9$ & 340.0 & 0.&7910 & 0.0218 \\
2010 JO$_{179}$ & 375.0 & 1.&7194 & 0.0635 \\
\\
2013 FY$_{27}$ & 362.5 & 1.&1722 & 0.0391 \\
2014 EZ$_{51}$ & 443.0 & 1.&6585 & 0.1012 \\
2010 RF$_{43}$ & 305.0 & 1.&1458 & 0.0227 \\
2003 OP$_{32}$ & 313.0 & 1.&3422 & 0.0288 \\
2012 VP$_{113}$ & 250.0 & 1.&0785 & 0.0118 \\
2010 KZ$_{39}$ & 260.0 & 1.&3606 & 0.0167 \\
2014 WK$_{509}$ & 218.5 & 1.&4378 & 0.0105 \\
\enddata

\end{deluxetable}

Table \ref{tbl:allDPs} list the physical parameters of the 34 largest TNOs in the Solar System observed as of February 2018 (the objects in the Minor Planet Center database, MPC, with absolute magnitude $H_V\le 4$); this set includes the 4 TNOs currently classified as dwarf planets by the IAU (Pluto, Eris, Haumea, and Makemake); however, for simplicity we call all of the large TNOs used in this study dwarf planets (DPs), regardless of their actual shapes or official classifications.

Note that only the 13 firsts objects in Table \ref{tbl:allDPs} have reliable measurements of their masses and densities (from Eris to 2003 AZ$_{84}$), the next 14 objects have only good estimations of their sizes (from Sedna to 2010 JO$_{179}$), while the last 7 count only with a measurement of their absolute magnitudes (from 2013 FY$_{27}$ to 2014 WK$_{509}$). Our procedure to assign masses and densities to the 21 objects without known data is described in detail in Appendix \ref{apdx:dps}.

We consider these 34 objects our DP population, similarly to what we have done in previous works when exploring the secular effect of DPs over cold debris disks \citep{Munoz15,Munoz17,Munoz18}. The initial conditions in heliocentric Cartesian elements (positions and velocities) for all of the massive objects in our simulations were retrieved from the JPL Horizons system for the Julian day 2458176.5, corresponding to February 27, 2018.

\begin{figure*}[htp]

  \centering
  \subfloat[Mass {\it vs.} semimajor axis for all the DPs in the simulations.]{\label{savsm}
  \includegraphics[width=.49\textwidth]{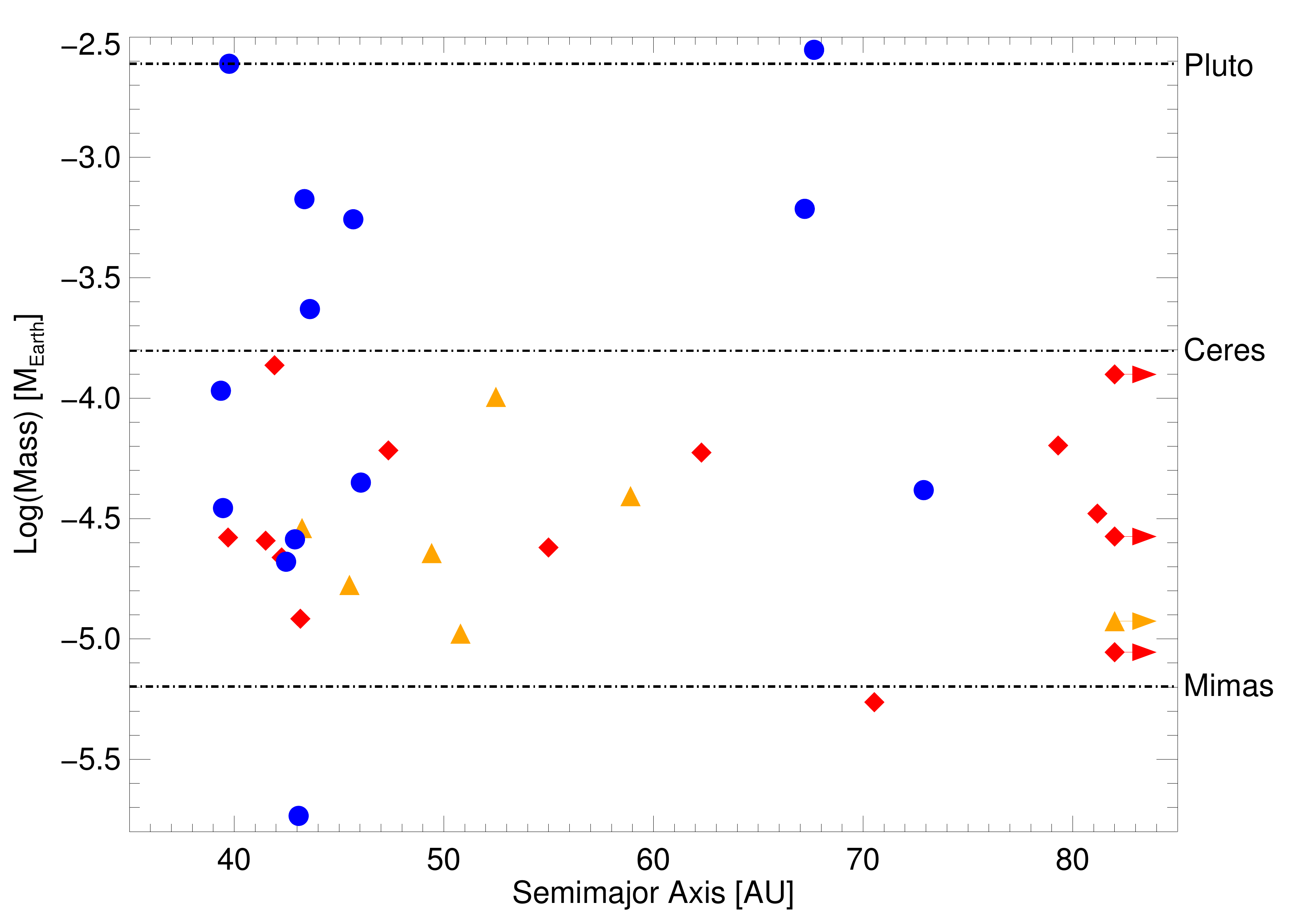}}
  \subfloat[Mass {\it vs.} perihelion distance for all the DPs in the simulations.]{\label{sqvsm}
  \includegraphics[width=.49\textwidth]{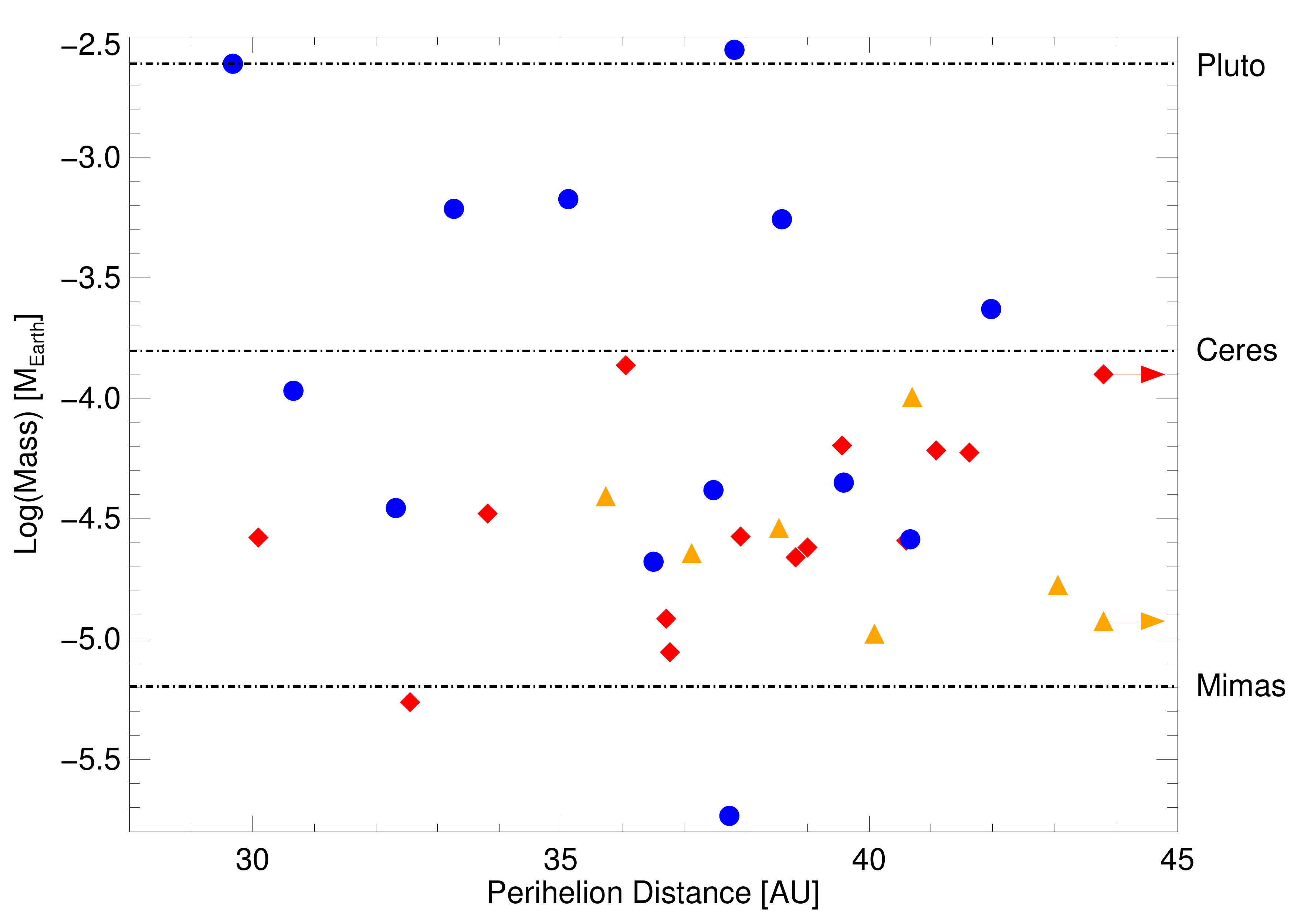}}

\caption{Distribution of DPs against their mass in M$_\oplus$. The blue circles in both panels stand for objects with known density and radius; red diamonds are for objects with known radius and random density; finally, orange triangles are for objects with known absolute magnitude but random albedo and random density. Symbols with arrows stand for objects with semimajor axes/perihelion distances larger than the scale of the figure. In panel \ref{savsm}, these are Sedna ($a\sim 484$ au), 2012 VP$_{113}$ ($a\sim 256$ au), 2005 QU$_{182}$ ($a\sim 112$ au), and 2014 UZ$_{224}$ ($a\sim 108$ au); while in panel \ref{sqvsm}, they are Sedna ($q\sim 76$ au) and 2012 VP$_{113}$ ($q\sim 80$ au).\label{fig:avsm}}

\end{figure*}

Fig. \ref{fig:avsm} shows the distribution of mass {\it vs.} semimajor axis, $a$, and perihelion distance, $q$, of the 34 DPs; the values for the mass of Pluto, Ceres, and Mimas are shown for reference. Note that, although 2002 TX$_{300}$ is much less massive than Mimas (blue circle below the Mimas line), we choose to include it in our set since, despite its small size, it has a measured density.

\subsection{Initial conditions for the Kuiper belt disk particles.}

\begin{figure}
\plotone{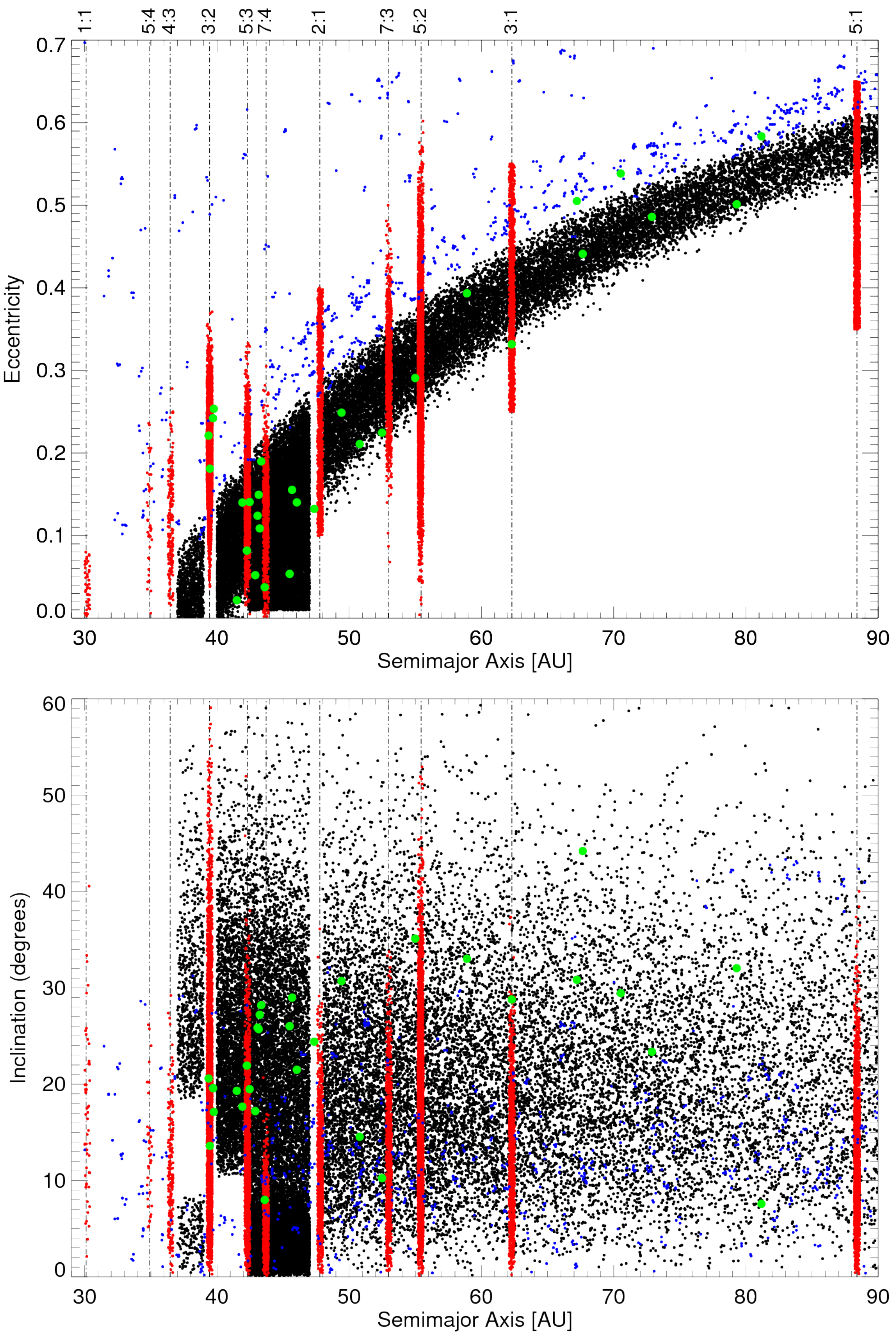}
\caption{Initial conditions of the particles in the L7 model as well as the DPs used in this study. Here we present the initial distribution of particles in the Classical (black dots), Resonant (red dots), and Scattering (blue dots) populations of the L7 model as well as the initial conditions of the DPs (filled green circles). The upper panel shows the semimajor axis vs. eccentricity phase-space plane; the lower panel shows the semimajor axis vs. inclination phase-space plane.\label{fig:l7model}}
\end{figure}

For the massless disk particles that integrate the population of cometary nuclei in the Kuiper belt, we used the L7 synthetic model described by \cite{Petit11,Gladman12}, which was made publicly available by the Canada France Ecliptic Plane Survey (CFEPS) team\footnote{\url{http://www.cfeps.net/?page_id=105}}. 
This is a model based on the CFEPS survey aiming to represent the true, unbiased orbital distribution of Kuiper belt objects down to an absolute magnitude of $H_g\le 8.5$. The model comprises: $50\,975$ classical particles (representing the cold, hot, and kernel populations of the classical KB), $13\,450$ resonant particles populating 11 mean motion resonances (MMRs) with Neptune, up to 4th order, and $1\,612$ particles which represent the scattering population.

In Fig. \ref{fig:l7model} we plot the distribution, in the $a$ {\it vs.} $e$ and $a$ {\it vs.} $i$ planes, of all the particles of the L7 synthetic model. Note that we do not make a distinction between classical and detached particles. This model suits the needs for our simulations, given that it represents a plausible distribution of objects in the Kuiper belt, while at the same time it provides a large enough population from which to draw statistically significant conclusions.

The implementation and use in our simulations of the classical and scattering populations in the L7 model is straightforward, as the initial conditions in osculating orbital elements are given by the CFEPS team. However, the resonant population requires a more careful treatment: we intend to use a population of particles well trapped in resonances; this to avoid any contamination introduced by the short-term stability of particles that are not protected by the resonant mechanism, despite having a semimajor axis similar to the one of the MMR; such particles, if used, would contaminate the rate of escapees from the Kuiper belt towards the Neptune crossing region that we are interested in measuring, falsely increasing the rate at which the secular perturbations from DPs are able to destabilize the objects of the KB.

In order to make use only of the actually resonant particles, we ran 10 Myr long simulations including only the 4 giant planets and the resonant population of the L7 model. In Section \ref{ss:resshort}, we summarize the results from this simulation, providing some statistics of the resonant particles. For the full term simulations we make use of a sub-population of $8\,371$ particles of the L7 model with libration amplitudes below 175$^\circ$, in any of the 11 MMRs.

\subsection{Simulation Parameters}

We performed six simulations using the symplectic integrator included in the MERCURY package of \cite{Chambers99}. The initial time-step was set to 400 days, with an accuracy parameter of $10^{-10}$ for the Bulirsch-St\"oer integrator, which goes into action when particles come closer than 4 Hill radii around each massive particle (giant planets and DPs). All of the simulations included a central star of 1+$\epsilon$ M$_\odot$ mass, where $\epsilon$ is the summed mass of the interior planets plus the Moon and Ceres. Three of the simulations included only the 4 giant planets as massive objects and the other three included the four giant planets and the 34 DPs as massive bodies, as well as the test particles in the Kuiper belt, with separate simulations for the classical ($50\,975$ particles), resonant ($8\,371$ particles), and scattering ($1\,612$ particles) populations.

The simulations were 1 Gyr each. We have used this integration time as a compromise between giving enough time for the secular effects to take hold while maintaining sensible CPU time requirements. Also, these simulations are designed to represent (more or less) the current behavior of the Solar System; during its formation the inner Solar System must have been a much more violent place, while in the far future less cometary nucleii are expected to remain and thus less new comets are expected to be injected. 

Note that one additional simulation used to characterize the resonant population is only 10 Myr long and includes only the 4 giant planets as massive objects, with the entire $13\,450$ set of resonant particles from the L7 model.

\subsection{Characterization of the resonant population of the L7 model}
\label{ss:resshort}

Depending on their angular elements, particles with semimajor axes with very close values to the nominal location of Neptune's MMRs, may or may not be librating in resonance with the planet. Particles actually trapped in MMRs with Neptune should be librating with amplitudes below 180$^\circ$. Those particles are protected from getting close enough to the giant as to be scattered off by it; thus they could, in principle, be long-term stable. This is of course not always the case, as chaotic perturbations can result in the stirring of eccentricities and in an increasing of the librating amplitude, which finally leads particles into encountering the giant. In any case, it is preferable to only make use of initially librating particles in the long-term simulations, in order to avoid a source of contamination coming from non-resonant particles that may be short-lived, even when they are located close to the resonance, but not protected by it.

In Fig. \ref{fig:l7model} the initial distribution of particles in the resonant population of the L7 synthetic model is shown by red dots. The resonant particles populate 11 MMRs with Neptune, from zero to fourth order, which are labeled at the top of the Figure. Those resonances are the following: 1:1, 2:1, 3:2, 4:3, 5:4, 3:1, 5:3, 5:2, 7:4, 5:1, and 7:3. After 10 Myr of dynamical evolution under the influence of the four giant planets, out of the $13\,450$ particles, initially in the 11 MMrs: 24 were ejected from the system (when their $a$ grew larger than 1000 au), 4 collided with the Sun or a planet, and from the remaining, we found that $8\,371$ particles librate during more than half of the integration with an amplitude below $175^\circ$, while the remaining $5\,051$ particles were non-resonant. We therefore define our resonant population, for the long term integrations of the next Section, as the subset of $8\,371$ librating particles.

\begin{deluxetable}{lccc}
\tabletypesize{\scriptsize}
\tablecaption{Characterization of the resonant particles in the L7 model.\label{tbl:resnum}}
\tablewidth{0pt}
\tablehead{
\colhead{MMR} & \colhead{Total number} & \colhead{Librating} & \colhead{Non-Resonant}  
}
\startdata
1:1 & 51   & 45   & 6   \\
2:1 & 867  & 655  & 212  \\
3:2 & 3327 & 3022 & 305  \\
4:3 & 195  & 150  & 45   \\ 
5:4 & 39   & 26   & 13   \\ 
3:1 & 942  & 500  & 442  \\
5:3 & 1259 & 716  & 543  \\
5:2 & 3178 & 1845 & 1333 \\
7:4 & 664  & 213  & 451  \\
5:1 & 2000 & 750  & 1250 \\
7:3 & 900  & 449  & 451  \\
\hline
sum & 13422 & 8371 & 5051 
\enddata


\end{deluxetable}

Table \ref{tbl:resnum} shows the number of particles initially in each resonance, as well as the number of particles that after 10 Myr were librating or turned out to be non-resonant. Note that for our long-term simulations, the 3:2 MMR contains the largest population, followed by the 5:2 MMR.

\section{Results and Discussion}
\label{sec:res}

\subsection{Evolution from the Kuiper Belt to Low Inclination Cometary Orbits}
\label{ssec:KB2JFC}

To analyze the evolution of particles, originally located in any of the families of the trans-Neptunian region (as given in the L7 model), we begin by defining three stages which any cometary nuclei, that effectively becomes a comet, is most likely to go through. These categories are as follows:
\begin{enumerate}
\item \emph{Crossers}: particles which perihelion, $q$, is smaller than $\anep + 2\sqrt{3}\rhn\approx 33$ au, where $\anep$ and $\rhn$ are the semimajor axis and Hill radius of Neptune, respectively (i.e. particles that cross into Neptune's region of influence). This limit is based on previous studies \citep{Gladman90,Bonsor12,Munoz18} which found that particles inside this region have a high probability of experiencing a close interaction with the giant planet, therefore jeopardizing their orbital stability. 
\item \emph{Nearly Interactive Particles} (NIPs): particles that get closer than one Hill radius to Neptune. These particles experience a close encounter with the giant and are strongly affected in their orbital motion. They can move inwards, towards Uranus, or outwards, away from the solar system.
\item \emph{Jupiter Family Comets} (JFCs). These are the ones that went inwards after interacting with Neptune, and then were handed down through the rest of the giant planets until interacting with Jupiter; each new step having the probability of passing the particle inwards or expelling it from the solar system. Once the cometary nuclei is close to Jupiter, the particle is defined as a JFC if its Tisserand parameter with respect to Jupiter, $\tj$, remains between 2 and 3 \citep[see, for example,][]{Levison97}.
\end{enumerate}

\begin{figure*}[htp]

  \centering
  \subfloat[Comets $a$ vs. $e$.]{\label{sfig:avse}
  \includegraphics[width=.48\textwidth]{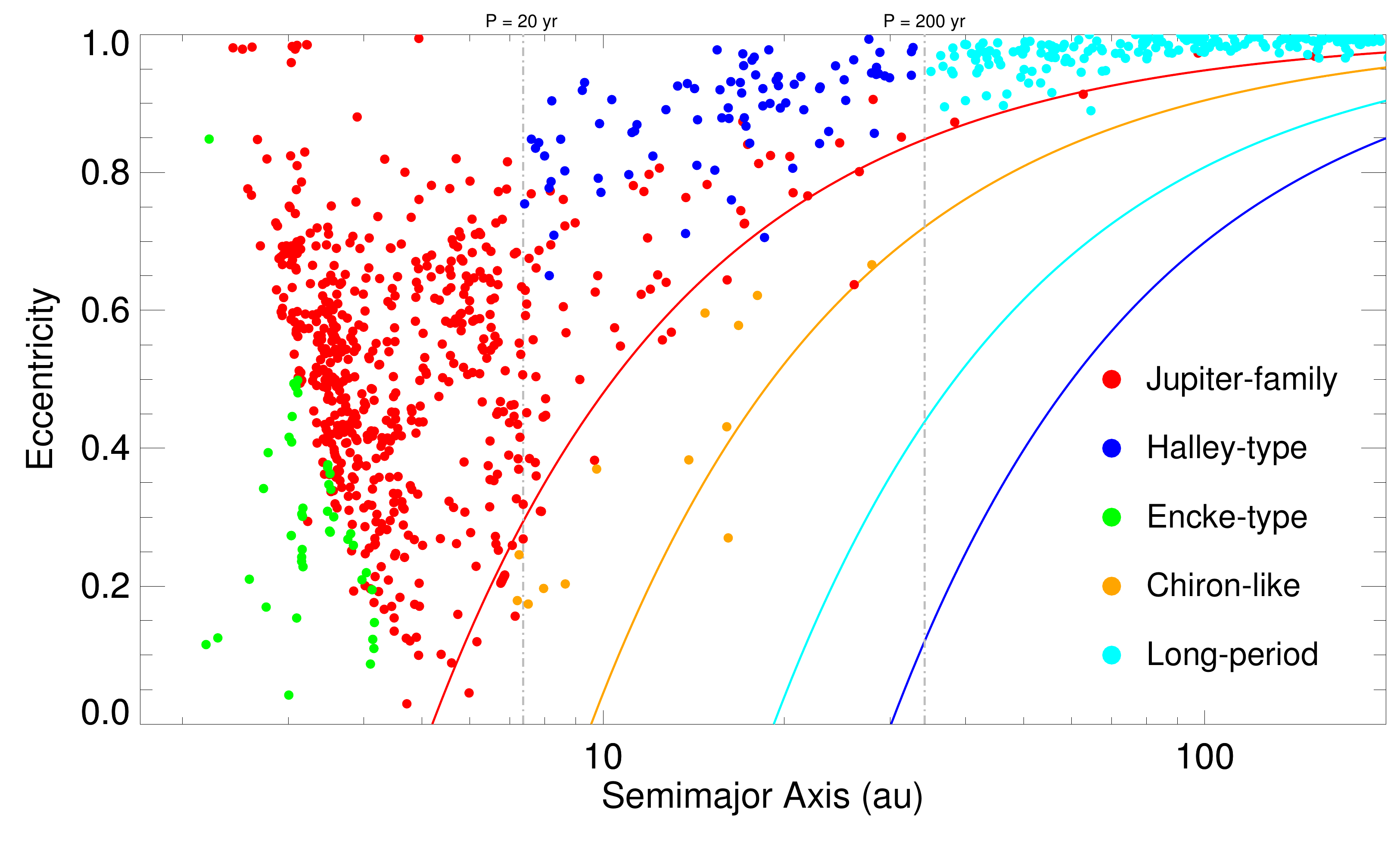}}
  \subfloat[Comets $a$ vs. $\tj$.]{\label{sfig:avstj}
  \includegraphics[width=.48\textwidth]{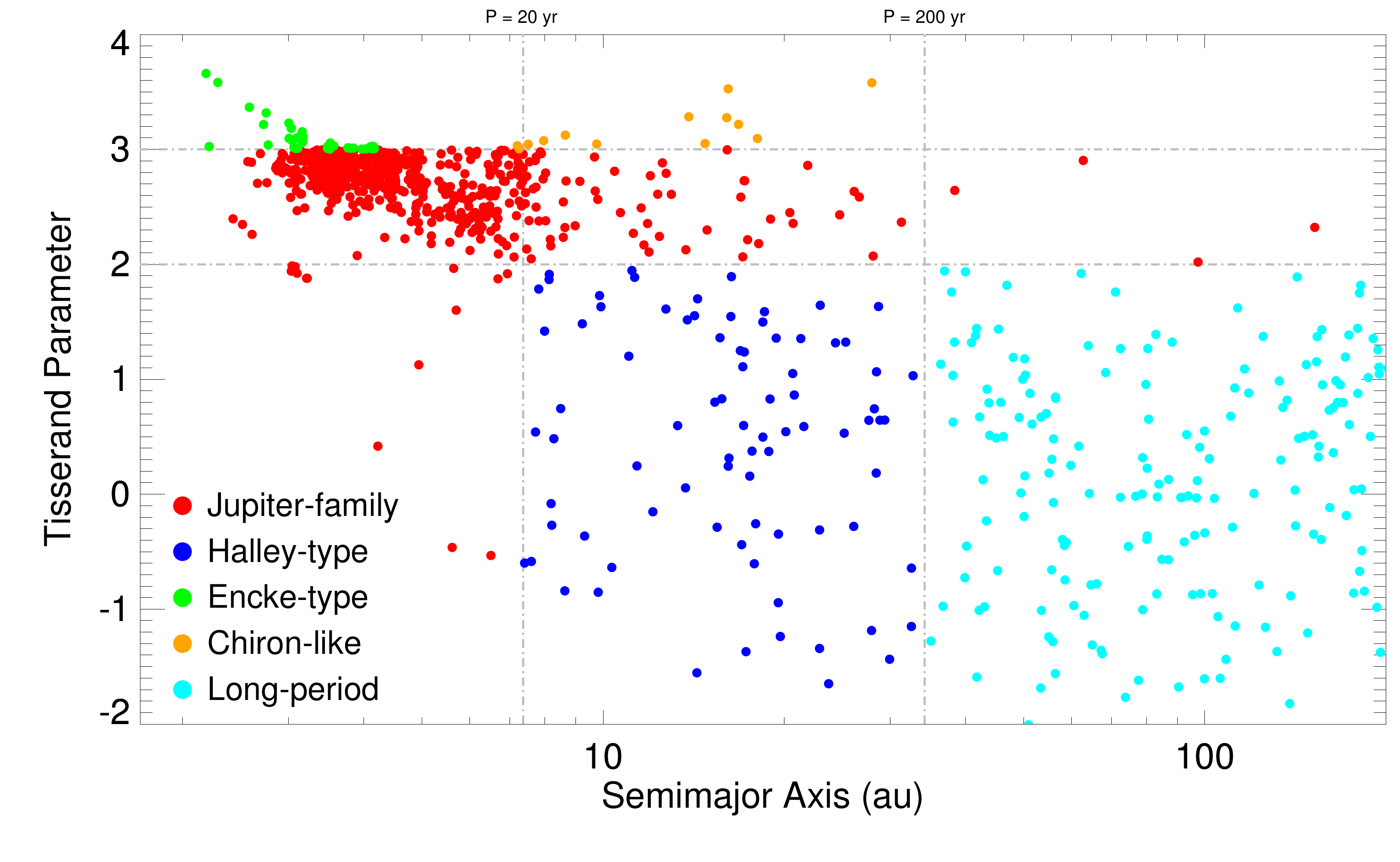}}

\caption{Distribution of known comets in phase space, as obtained from the JPL Small Body Search Engine (as of Oct. 2018). In total 826 short-period comets are shown (with $P<20$ yr), out of which 674 are JFCs as defined by \citet{Levison97}: $2<\tj<3$. We plotted JFCs as red circles and the other families are shown for context.\label{fig:comphsp}}

\end{figure*}

The Tisserand parameter is given by
\begin{equation}
\tj=\frac{\asj}{a}+2\sqrt{(1-e^2)\frac{a}{\asj}}\cos{i},
\end{equation}
where $\asj$ is Jupiter's semimajor axis, and $a$, $e$, and $i$ are the semimajor axis, eccentricity, and inclination of the particle. It has been shown that $\tj$ is a suitable parameter to differentiate between the Jupiter Family and Halley-type comets, both of which are considered SPCs (historically, those with period below 200 yr; the JFCs having periods below 20 yr); however, by only considering the period as a classifying criteria, one ends with a large overlap between both families, thus a dynamically motivated criterion such as $\tj$ turns out to be a better choice in defining families among the observed comets, with Halley-types having a $\tj<2$ \citep{Kresak72}. In Fig. \ref{fig:comphsp} we show the distribution of all the known comets as of October 2018 (obtained from the JPL Small-Body Database search Engine). 

We consider the perihelion, $q$, in order to count a particle in our simulations as a comet, assuming that the cometary nuclei would become visible the first time $q$ becomes less than 2.5 au (i.e, it would display cometary activity in the form of a coma and a tail); this is a common cut used in different works \citep[e.g.][]{Levison97,Nesvorny17} when attempting to compare the results from numerical simulations with the observed population of comets in the solar system.

From the simulations, we saved the information relative to close encounters of all particles below 4 Hill radius of any major body; from those files we can extract the time at which one particle first becomes a member of any of the 3 categories defined above. To measure the effect of DPs over the evolution of the Kuiper belt's cometary nuclei, we compare the cumulative fraction of particles for simulations with and without DPs; we do this separately for each population of the L7 model; i.e. Classical, Resonant, and Scattering. Please note that objects that start in the Classical (or Resonant) population that become crossers, NIPs, or JFCs will most likely be part of the Scattered population for a period of time, none the less we will consider them as members of the Classical (or Resonant) population, respectively, for the remaining of the paper. 

The crosser stage represents an ``initial'' step in the evolution towards the inner planetary system: after being perturbed inside the non planet-crossing region of the Kuiper belt, a particle enters a region where it can be strongly perturbed and its orbit significantly modified by Neptune. If a particle gets close enough to Neptune (a NIP), the interchange of energy and angular momentum with the giant will alter the orbit severely, sending the particle inwards or outwards. If inwards, the same interchange process can occur with the rest of the giant planets until the particle is stabilized in a cometary orbit by Jupiter, with a nearly constant Tisserand parameter between 2 and 3; the particle will remain in this region for at least a few thousands of years, being observable as a comet while it remains active. 

In what follows, we consider a JFC as a particle that satisfies the two above criteria ($2<\tj<3$ and $q<2.5$ au) for the first time during its orbital evolution. It is worth to note that the precision of our simulations to follow orbits after they first become JFCs is limited (this due to: our integration timesteps, our output cadence, the absence of inner planets, and the lack of a model for the outgassing of the comet); therefore we prefer to restrict ourselves up to this instant in the history of the cometary nuclei only, when accounting for the effect of DPs on the injection rate of such comets.

The separated results for each of the L7 model populations are presented below.

\subsection{Long-term Evolution of the Resonant Populations}
\begin{figure*}[htp]

  \centering
  \subfloat[Crossers vs. time.]{\label{cvstRes}
  \includegraphics[width=.32\textwidth]{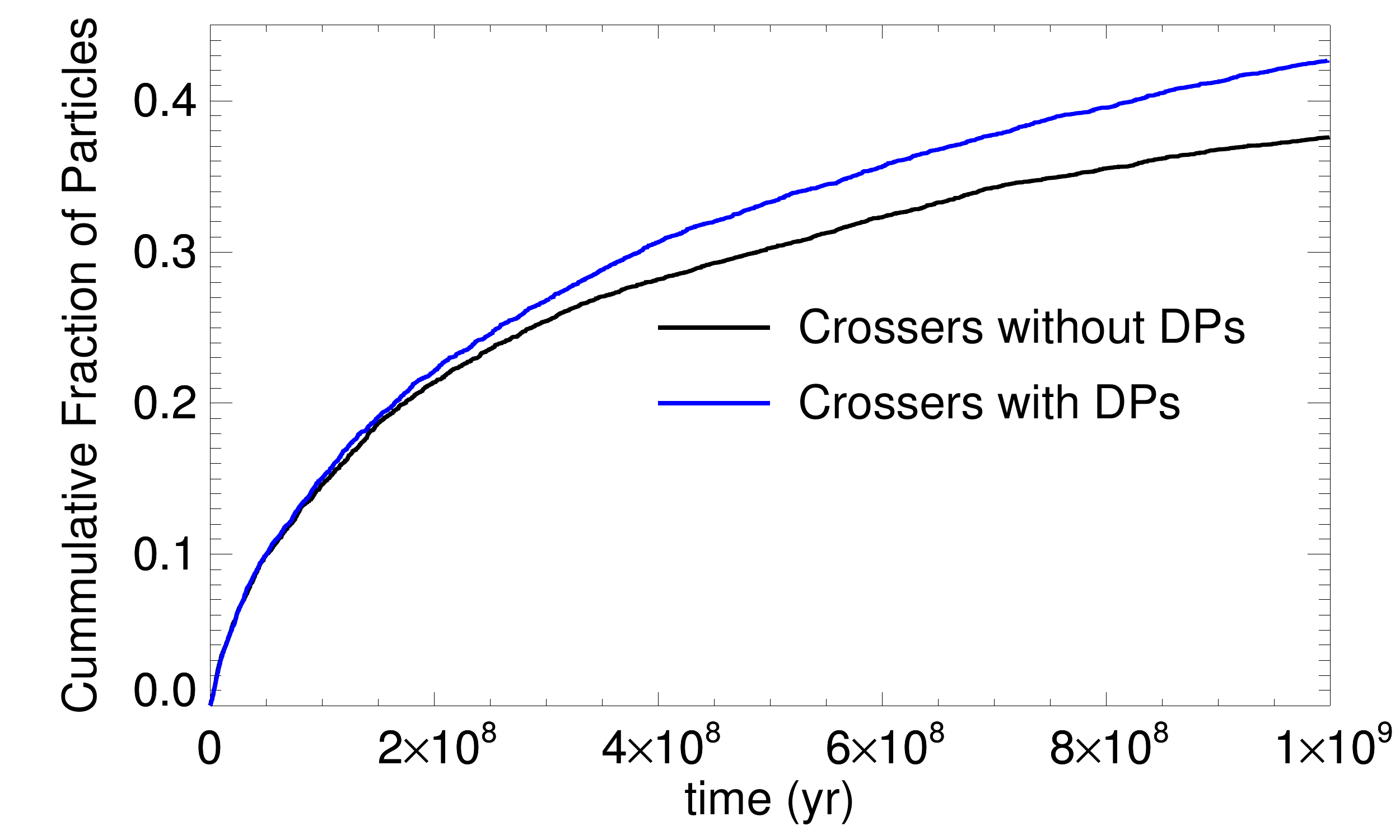}}
  \subfloat[NIPs vs. time.]{\label{nipsRes}
  \includegraphics[width=.32\textwidth]{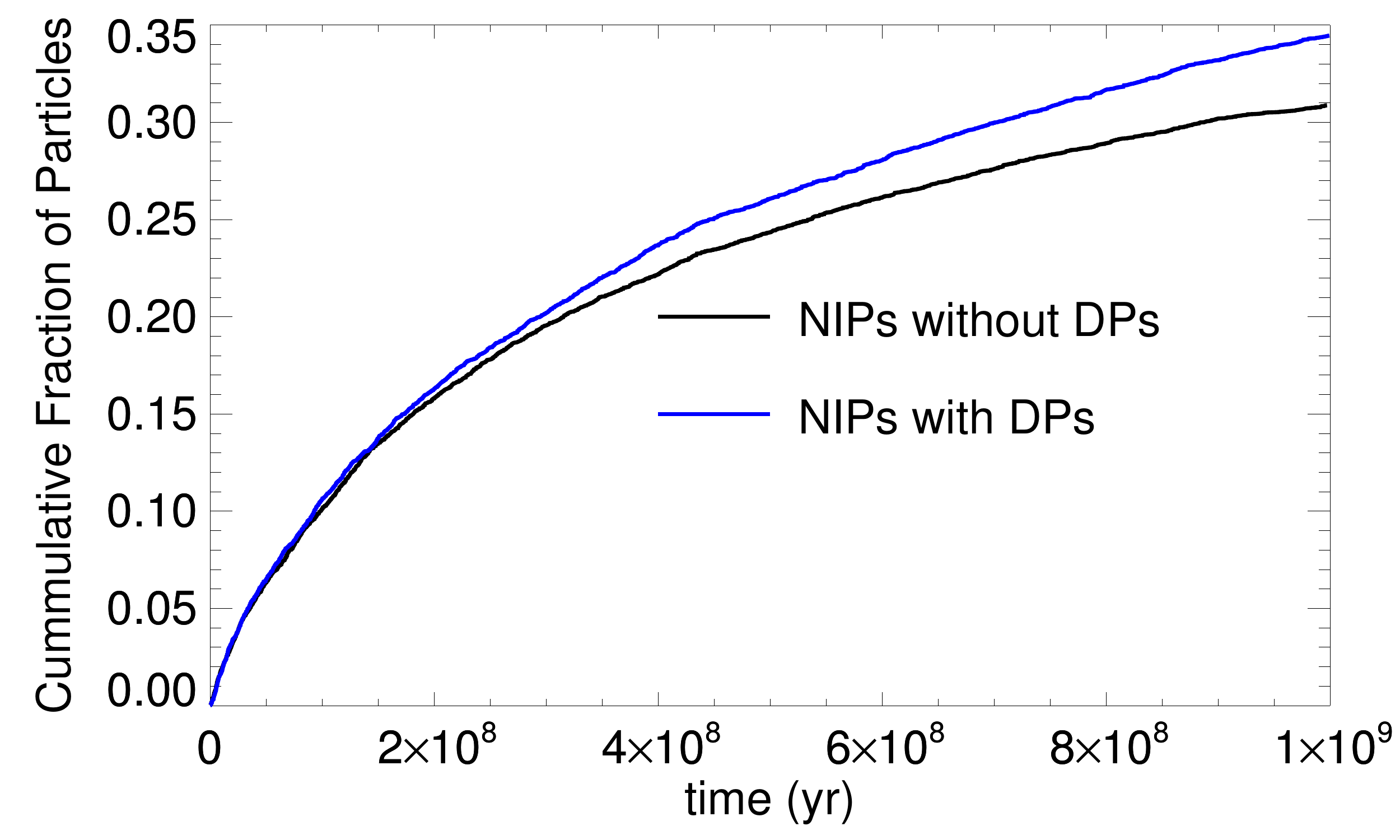}}
  \subfloat[JFCs vs. time]{\label{jfcRes}
  \includegraphics[width=.32\textwidth]{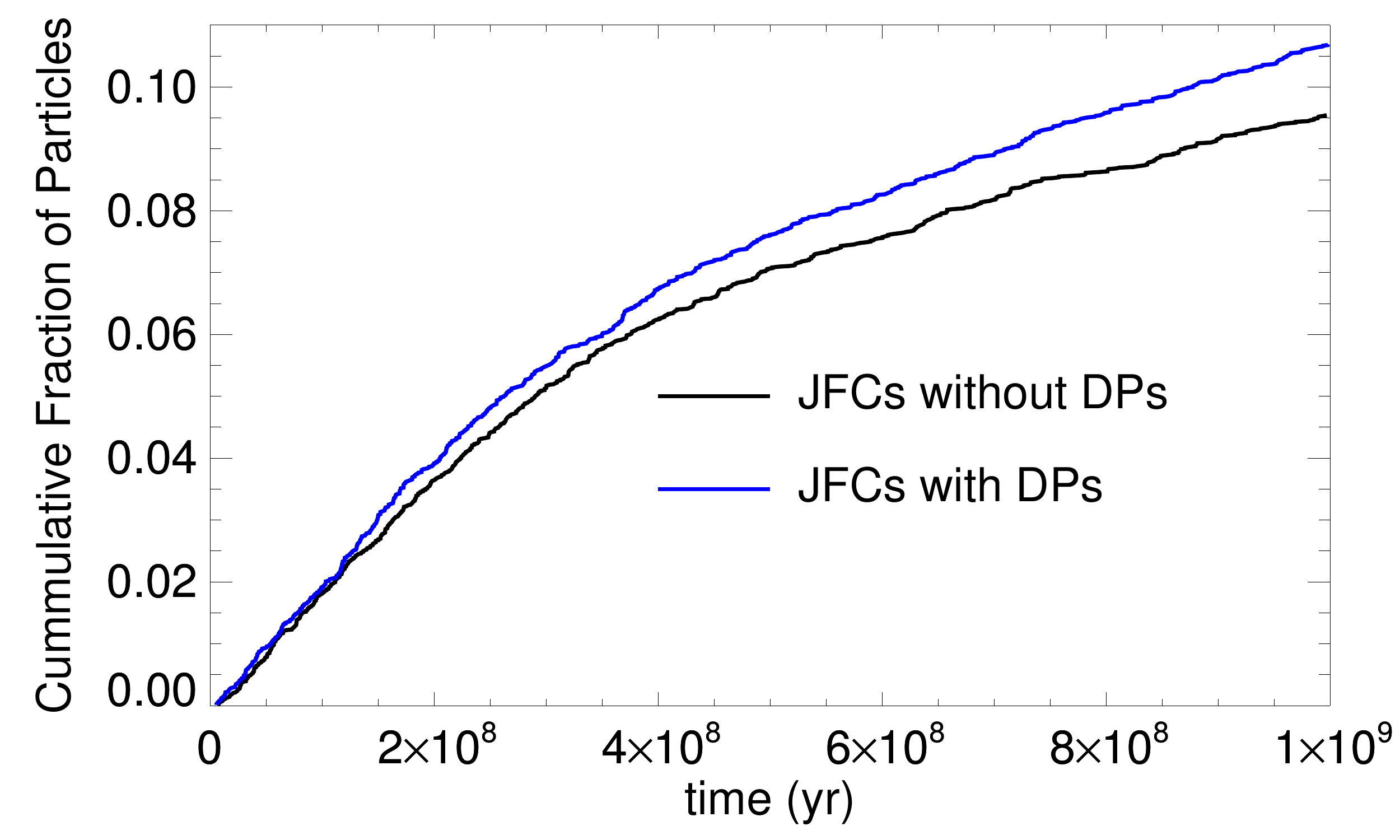}}

\caption{Cummulative fractions of particles that become crossers, NIPs, and JFCs coming from the resonant population. Both, in the presence (blue lines) and absence (black lines) of DPs.\label{fig:fracsRes}}

\end{figure*}

We performed two simulations (1 Gyr long) with the $8371$ resonant particles identified in Section \ref{ss:resshort}, this population represents 13.7\% of all test particles used in this work. The first simulation included only the 4 giant planets as massive objects, while the second also included the 34 DPs of Tables \ref{tbl:knownDPs}, \ref{tbl:unknownDPs}, and \ref{tbl:albDPs}.

The cummulative fraction of particles that become crossers, NIPs, and JFCs are shown in Figure \ref{fig:fracsRes}. The black lines are the cummulative fractions from the simulation without DPs, while the blue lines are from the simulation with DPs.

We observe a significant increase in the number of JFCs in the simulation with DPs, confirming that the interactions with the DPs have indeed a significant secular effect in the evolution of the resonant populations on Gyr time-scales. Actually, the three families of particles experiment an increase in the total number of their respective members. This effect has a delay of $\sim 150$ Myr in all cases, showing the secular nature of this phenomenology, after this time, the blue curves remain above the black curves, revealing that the perturbations produced by DPs are enough to increase the population of unstable resonant particles in the Kuiper belt. 

The numbers of crossers, NIPs, and JFCs at the end of the simulation without DPs are 3147, 2585, and 799, respectively; while, for the simulation with DPs, those numbers are 3571, 2885, and 895. The effective increases of $\sim$ 13.5\%, 11.6\%, and 12.0\%, respectively, are due to DPs. We can also observe that in both simulations, around 80\% of the crosser particles become NIPs, and from those, around 30\% become JFCs (25\% of the crossers become JFCs). Those last fractions are independent of the presence of DPs, as one would expect, since after the particles reach the dominion of the giant planets, their outcome will mostly depend on the interchanges of energy and angular momentum with Neptune and the other giant planets, compared to which the perturbations from DPs are negligible. Previous works have also found similar fractions for the particles that become visible JFCs after first encountering Neptune \citep{Levison97,Volk13}. 

\subsubsection{The Contribution of Individual Resonances to the Population of JFCs}

\begin{figure}
\plotone{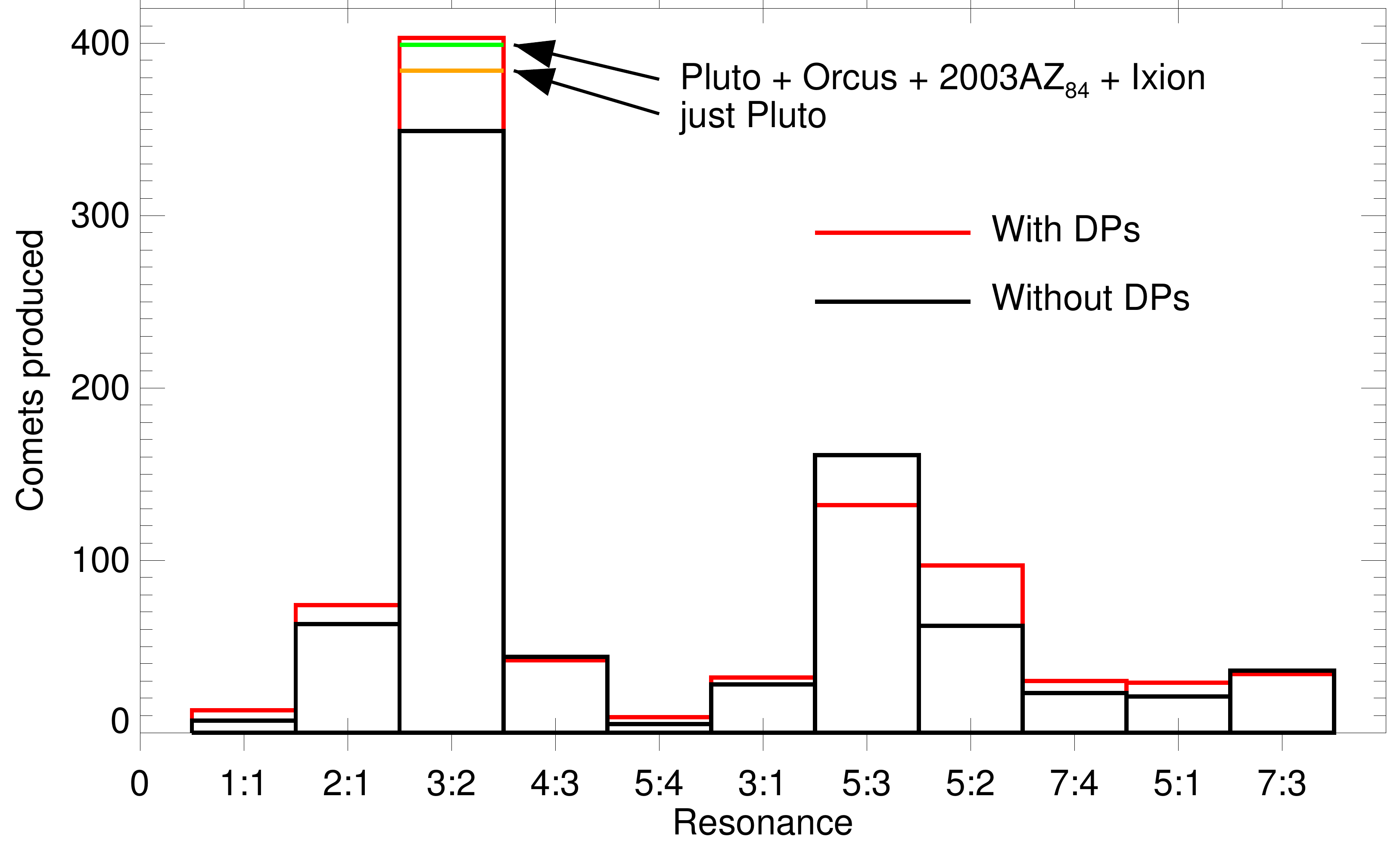}
\caption{Number of new visible JFCs produced by each resonance in simulations with DPs (red line) and without DPs (black line). Also shown are the numbers of JFCs produced in the two long-term simulations of the 3:2 population: with Pluto as massive object (orange line), and with the four massive Plutinos (green line). \label{fig:hisRes}}
\end{figure}

In Section \ref{ss:resshort}, we define the resonant population on the basis of their libration in each of the 11 MMRs described in the L7 model. We found that only little more than 60\% of the original particles --- of those defined as resonant in the L7 model --- actually librate in each of the MMRs with Neptune, while a little less than 40\% are non-resonant. The MMRs are not homogeneously populated nor do they have a homogeneous librating fraction; in fact, the 3:2 and 5:2 populations alone comprise almost 50\% of the original population in the L7 model (see Table \ref{tbl:resnum}), while, after our characterization, they comprise around 60\% of our librating sample. Therefore, it would be expected that the majority of new JFCs, that escaped from the resonant regions, would actually come from the 3:2 and 5:2 MMRs.

Fig. \ref{fig:hisRes} shows the number of new visible JFCs proceeding from each MMR with Neptune for the simulations with and without DPs. The black columns show the number of comets from each MMR for the simulation without DPs, while the red columns stand for the simulation including DPs. A great deal of new JFCs come originally from only 2 resonances: the 3:2 (43.7\% of all JFCs without DPs, and 45.0\% when DPs are present) and the 5:3 (20.1\% without DPs, 14.7\% with DPs). 

Interestingly, the behavior of the 5:3 MMR (and to a lesser extent the 7:3 and the 4:3 MMRs) is contrary to the rest of the MMRs: when the DPs are included in the simulations the number of JFCs produced by most MMRs increases, while for the 5:3 MMR this number decreases by 18\%. Overall, the total number of JFCs produced by MMRs increases by 12\% when DPs are present.

Given its contribution of up to 45\% of all the JFCs coming from MMRs, the Plutino population deserves an independent analysis. We observe an increment of around 15\% in the number of comets produced by the 3:2 resonance as a result of the inclusion of DPs in the simulation; this is the second largest percentile increment for any of the MMRs analyzed (second only to the 56\% increase in the 5:2 MMR) but the largest in actual numbers (54 additional comets). We recall that our population of DPs includes four Plutinos: Pluto, Orcus, 2003 AZ$_{84}$, and Ixion. To understand the effect of these objects on the diffusion of the whole Plutino population, we perform two additional 1 Gyr simulations including only the 3022 massless particles of the 3:2 MMR, with the giant planets and Pluto as big objects in the first case, and the giant planets and Pluto, Orcus, 2003 AZ$_{84}$, and Ixion as large objects in the second case.

For the two above simulations, we present in Fig. \ref{fig:hisRes} the number of JFCs produced in each case. With only Pluto, the number of JFCs produced is 384, an effective increase of 10\% with respect to the 349 JFCs obtained when no DPs are included in the simulations. On the other hand, when the four massive Plutinos are included, there are 399 JFCs, nearly identical to the 403 JFCs obtained in the simulation that included all the 34 DPs.

In a previous work, \citet{Tiscareno09} found that Pluto has only a modest effect, decreasing in 3\% the mean particle lifetime in the 3:2 resonance, i.e., an expected decrease of 2\% in the number of particles remaining in resonance after 1 Gyr. Although our estimation on the significance of Pluto when producing JFCs may appear significantly larger than the \cite{Tiscareno09} estimate, in reality the fraction of particles remaining in the resonance is only reduced by around 8\% due to the presence of Pluto. This percentage increases slightly in the simulation that included the four massive Plutinos as well as in the one with the 34 DPs. 

\subsection{Long-term Evolution of the Classical Population}

\begin{figure*}[htp]

  \centering
  \subfloat[Crossers vs. time.]{\label{cvstClass}
  \includegraphics[width=.32\textwidth]{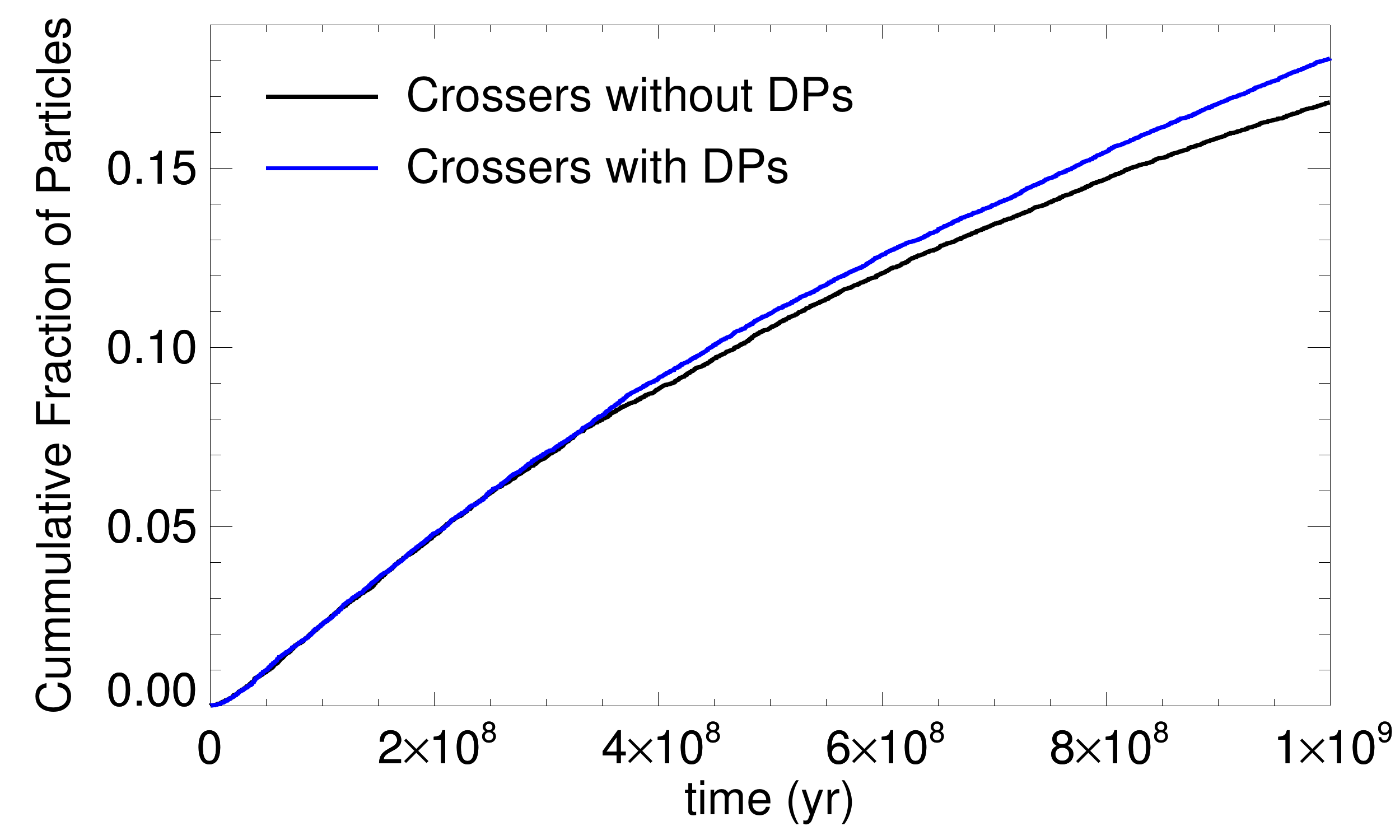}}
  \subfloat[NIPs vs. time.]{\label{nipsClass}
  \includegraphics[width=.32\textwidth]{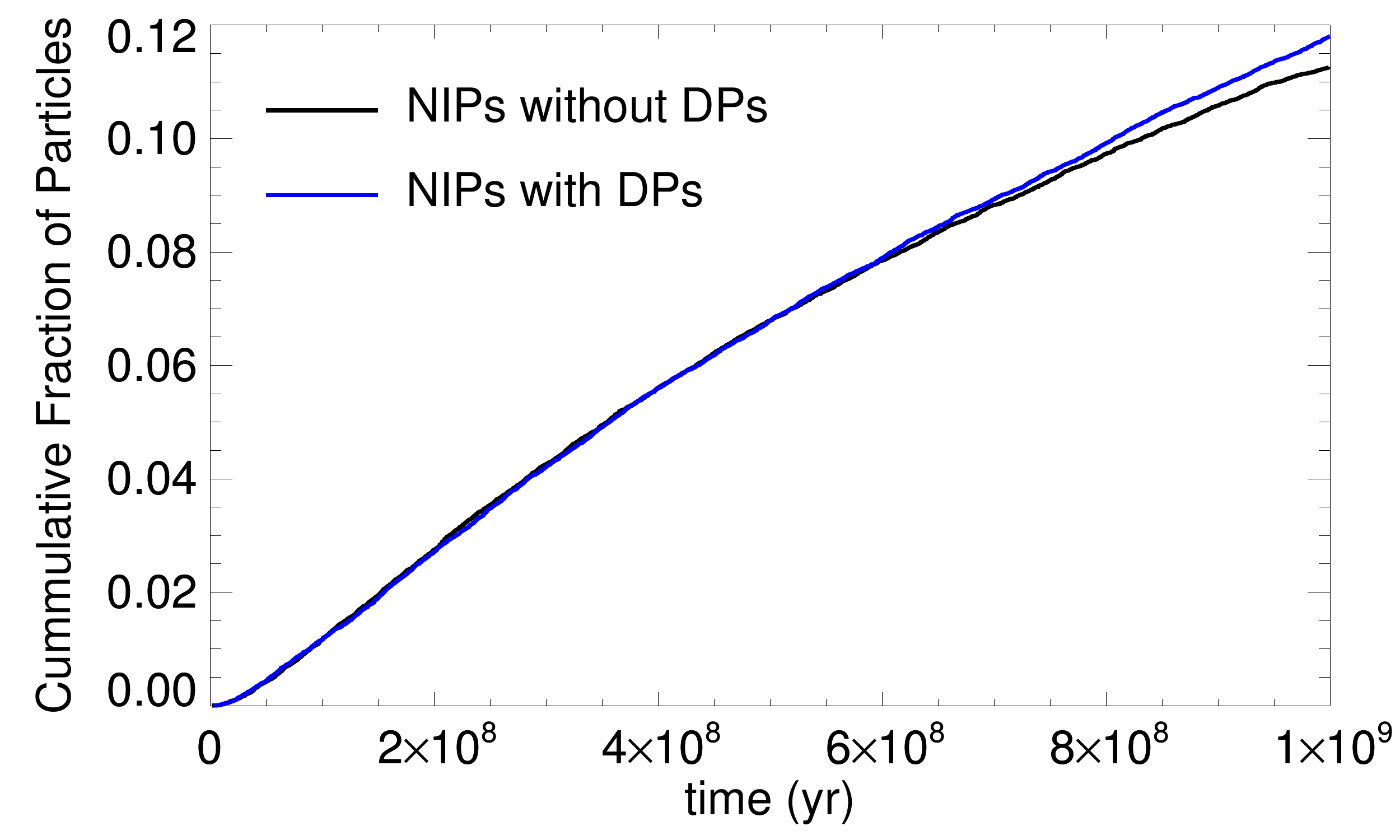}}
  \subfloat[JFCs vs. time]{\label{jfcClass}
  \includegraphics[width=.32\textwidth]{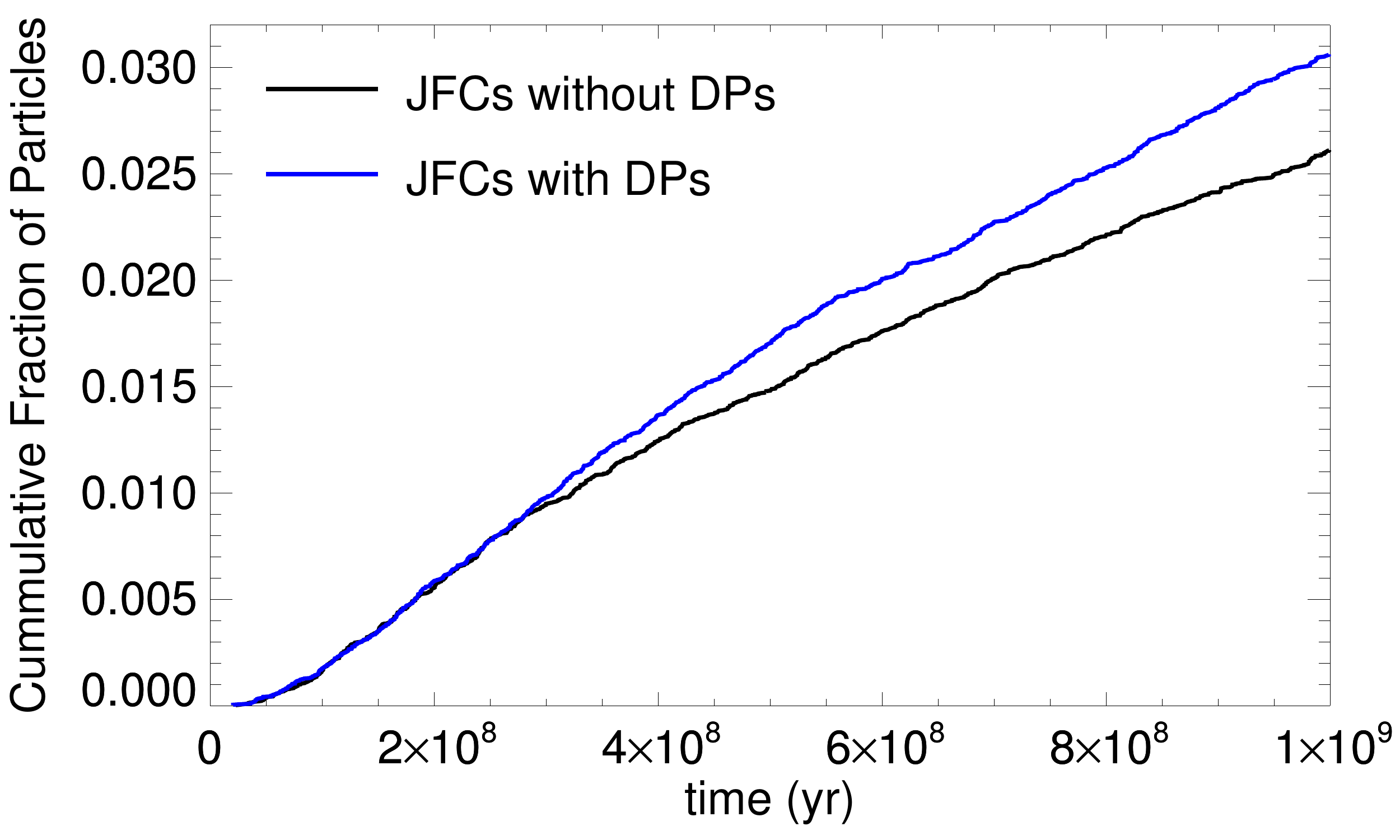}}

\caption{Cumulative fraction of particles that become crossers, NIPs, and JFCs coming from the classical population. Both, in the presence (blue lines) and absence (black lines) of DPs. \label{fig:fracsClass}}

\end{figure*}

The classical population is initially formed by 50975 particles with $a<500$ au and $q>33$ au. This population represents about 83.6\% of all the particles simulated in this work, thus it is expected to be the main source of JFCs. In Fig.~\ref{fig:fracsClass} we show the evolution of the fraction of particles that become crossers, NIPs, and JFCs in the simulations of the classical population, with DPs (blue lines) and without DPs (black lines).

In the same way as for the resonant population, we can see an increment in the fractional number of particles that become crossers, NIPs, and JFCs, due to the presence of DPs; thus reaffirming their significance, this time for the broader population of the classical Kuiper belt.

The numbers of crossers, NIPs, and JFCs obtained in the simulation without DPs are 8586, 5741, and 1332, respectively, while, in the simulations with DPs, these numbers become 9211, 6018, and 1561. The observable effect of DPs over the fraction of particles in each stage shows a delay of approximately 400 Myr for the crossers, almost 700 Myr for the NIPs, but only 300 Myr for the JFCs. The increase in the fraction of particles induced by DPs is 7.3\%, 4.8\%, and 17.2\% for each stage. For the classical population, without DPs, 66.8\% of crossers become NIPs, and from these, 23.2\% become JFCs (15.5\% of crossers become JFCs); on the other hand, when DPs are included, these percentages change slightly, with 65.3\% of the crossers becoming NIPs, and 25.9\% of those reaching the JFCs stage (in this case, 16.9\% of crossers become JFCs). These numbers show that, for the classical population, the added particles reaching the stage of NIPs due to perturbations from DPs, have a slightly larger probability of becoming JFCs. 

\subsection{Long-term Evolution of the Scattering Population}

\begin{figure*}[htp]

  \centering
  \subfloat[Crossers vs. time.]{\label{cvstSca}
  \includegraphics[width=.32\textwidth]{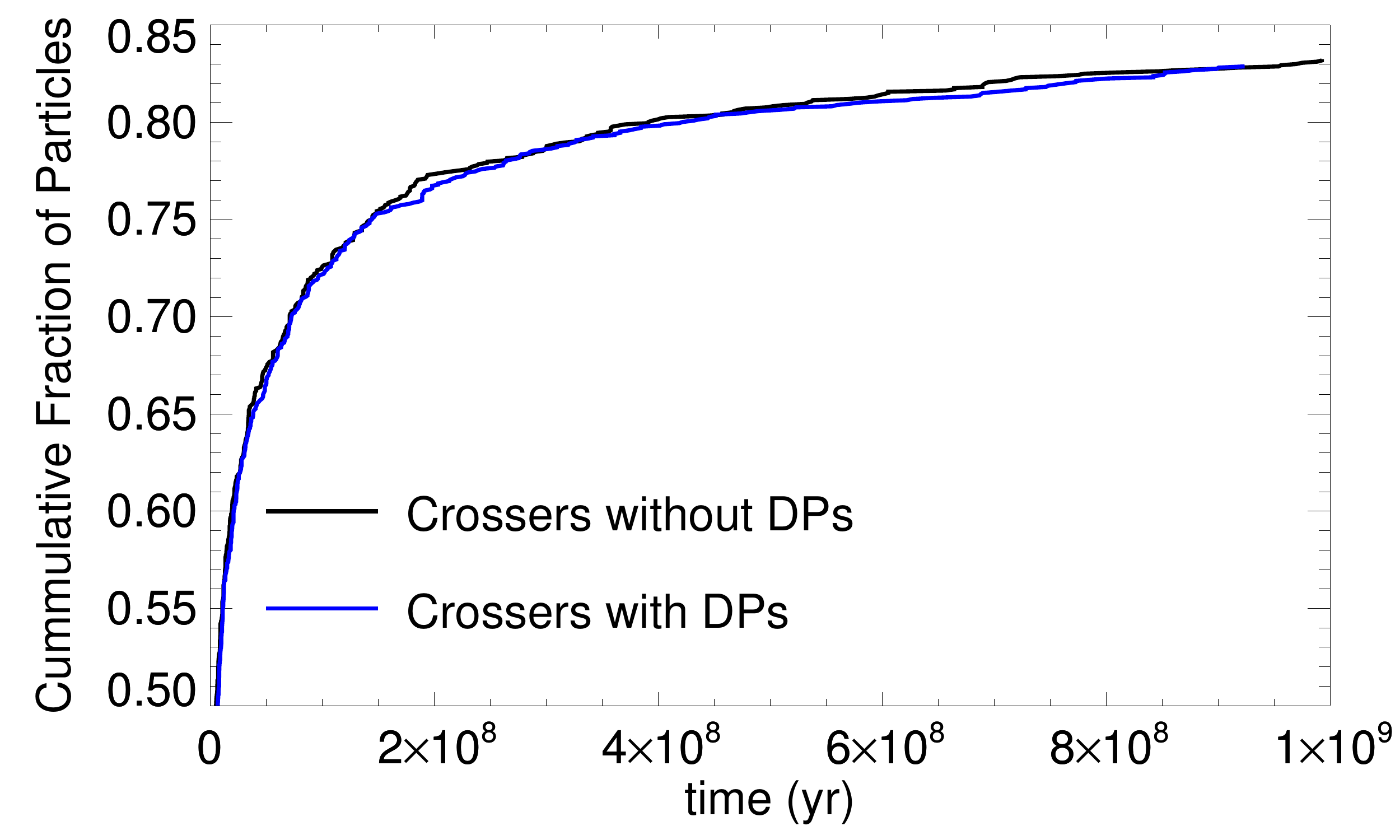}}
  \subfloat[NIPs vs. time.]{\label{nipsSca}
  \includegraphics[width=.32\textwidth]{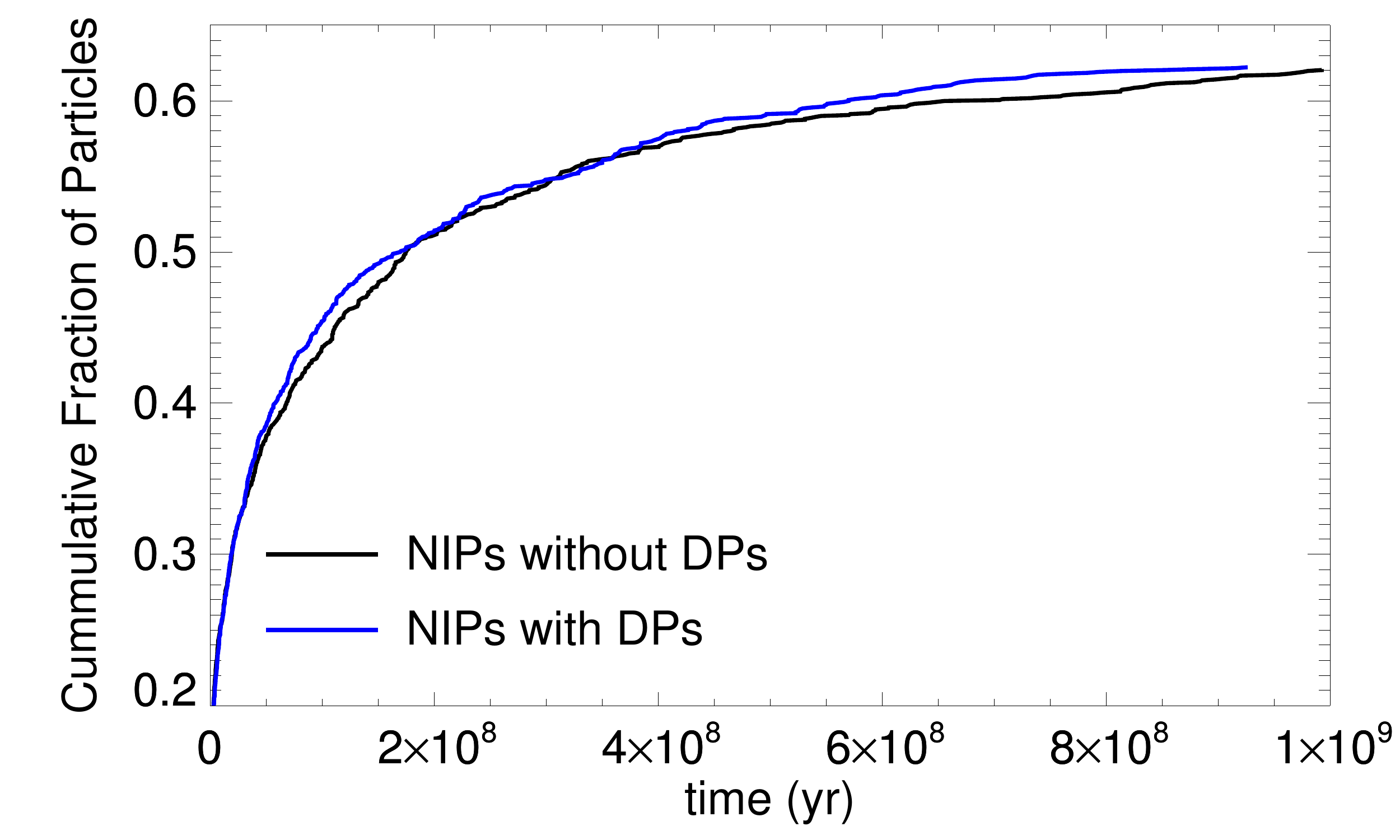}}
  \subfloat[JFCs vs. time.]{\label{jfcSca}
  \includegraphics[width=.32\textwidth]{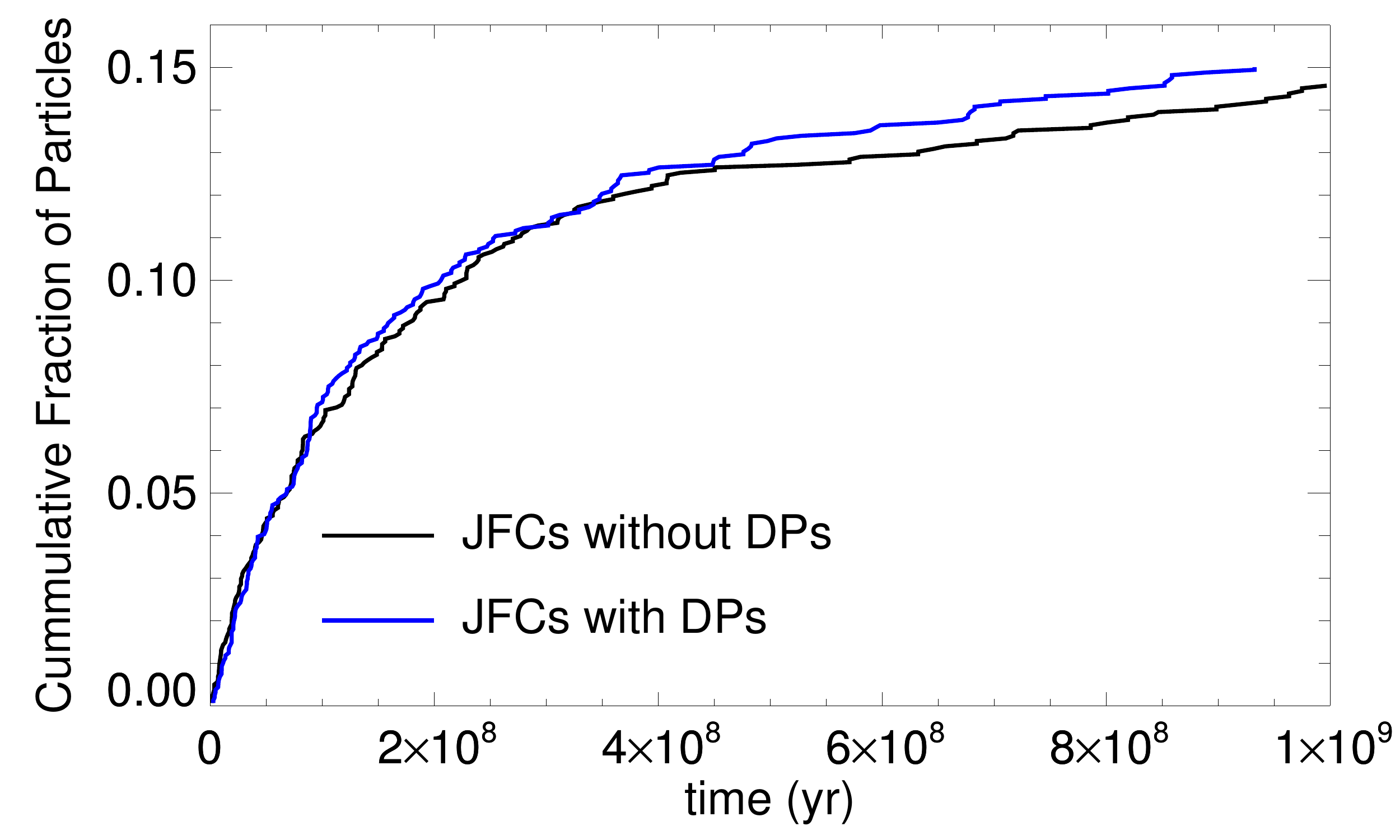}}

\caption{Cumulative fraction of particles that become crossers, NIPs, and JFCs coming from the scattering population. Both, in the presence (blue lines) and absence (black lines) of DPs.\label{fig:fracsSca}}

\end{figure*}

The scattering population in the L7 model is formed by only 1612 particles, which account for 2.7\% of all the particles simulated in this work. In Fig. \ref{fig:fracsSca} we show the evolution of the fractional number of particles for the crossers, NIPs, and JFCs categories in the simulations with DPs (in blue lines) and without DPs (in black lines). For this population, there is not an appreciable increase in the number of crossers and NIPs due to the presence of DPs, and only a slight increase in the production of JFCs is obtained when DPs are included.

The number of crossers, NIPs, and JFCs in the simulation without DPs are 1342, 1001, and 235, respectively. On the other hand, with DPs these numbers are 1336, 1003, and 242, respectively. In this case, it is not straightforward to link the slight increase of comets to the effect of DPs, since such small variations could result from statistical fluctuations expected from the 2 different simulations (yet the presence of DPs result in a 3\% increase in the total number of JFCs). 

Most of the scattering objects are destined to evolve in this track. Note that more than 80\% of the total population of scattering objects become crossers, while more than 60\% of the whole population become NIPs. Actually, for both simulations, around 75\% of crossers become NIPs, and from those, around 23\% become JFCs (around 17\% of crossers become JFCs). 

\subsection{Overall Dynamical Heating Produced by DPs on the TNO Cometary Populations}

\begin{figure}
\plotone{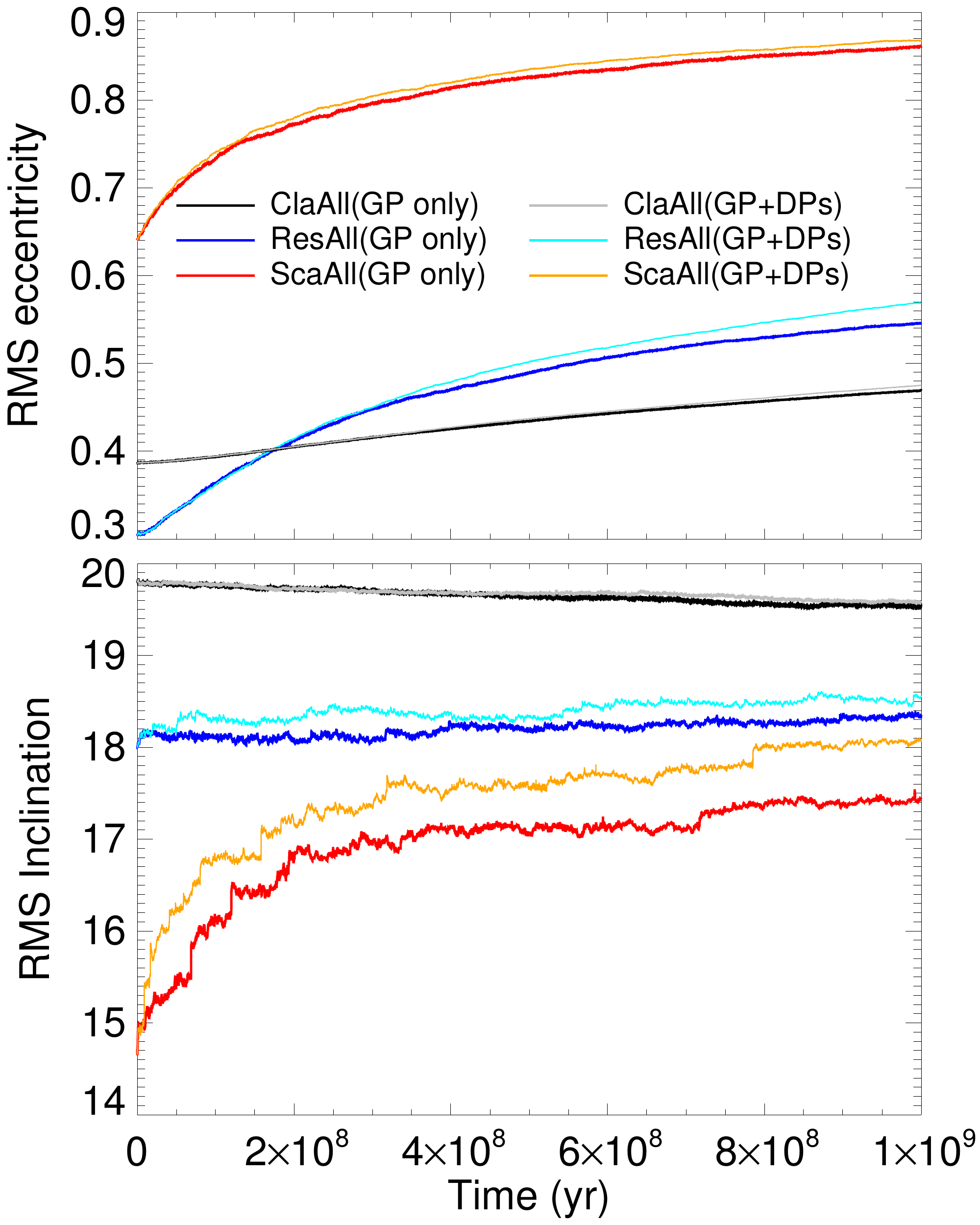}
\caption{Evolution of the rms eccentricity and inclination of all the particles in the three populations of the L7 model, as a function of time. The upper panel shows the evolution of the rms eccentricity for the Classical population (gray and black lines, with and without DPs, respectively), the Resonant population (cyan and blue lines, with and without DPs, respectively), and the Scattering population (orange and red lines, with and without DPs, respectively). The lower panel shows the evolution of the rms inclination for the same three populations, with and without DPs, with the same color code as the upper panel. \label{fig:rmsall}}
\end{figure}

\begin{figure}
\plotone{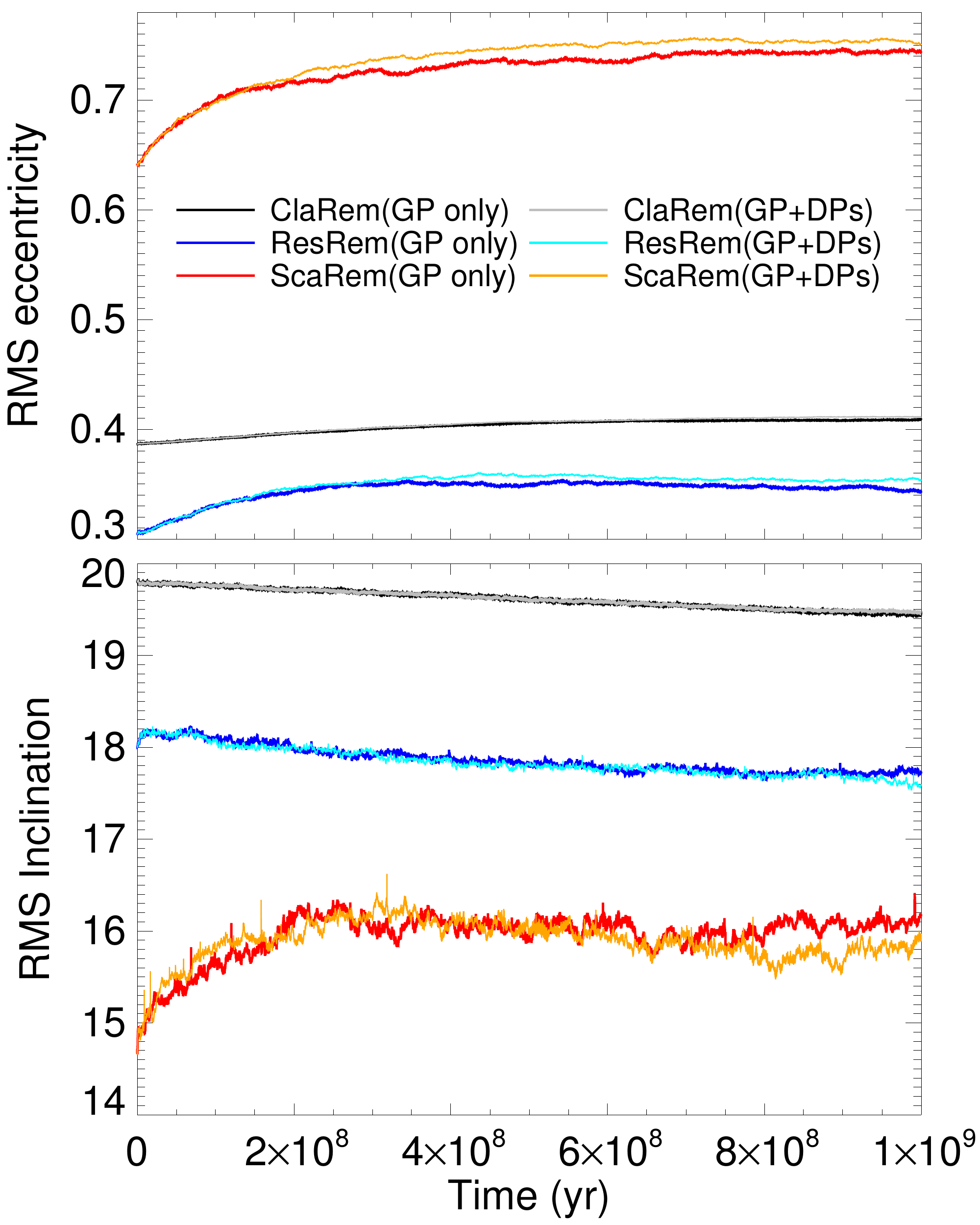}
\caption{Same as Figure \ref{fig:rmsall}, but only averaging the particles that remain in the simulation; that is, particles that have not collided nor reached a distance of 1000 au from the Sun; in other words, particles that would be observable in the Solar System. The upper panel shows the evolution of the rms eccentricity and the lower panel the evolution of the rms inclination. The color code is the same as in Fig. \ref{fig:rmsall}\label{fig:rmsrem}}
\end{figure}

\begin{figure}
\plotone{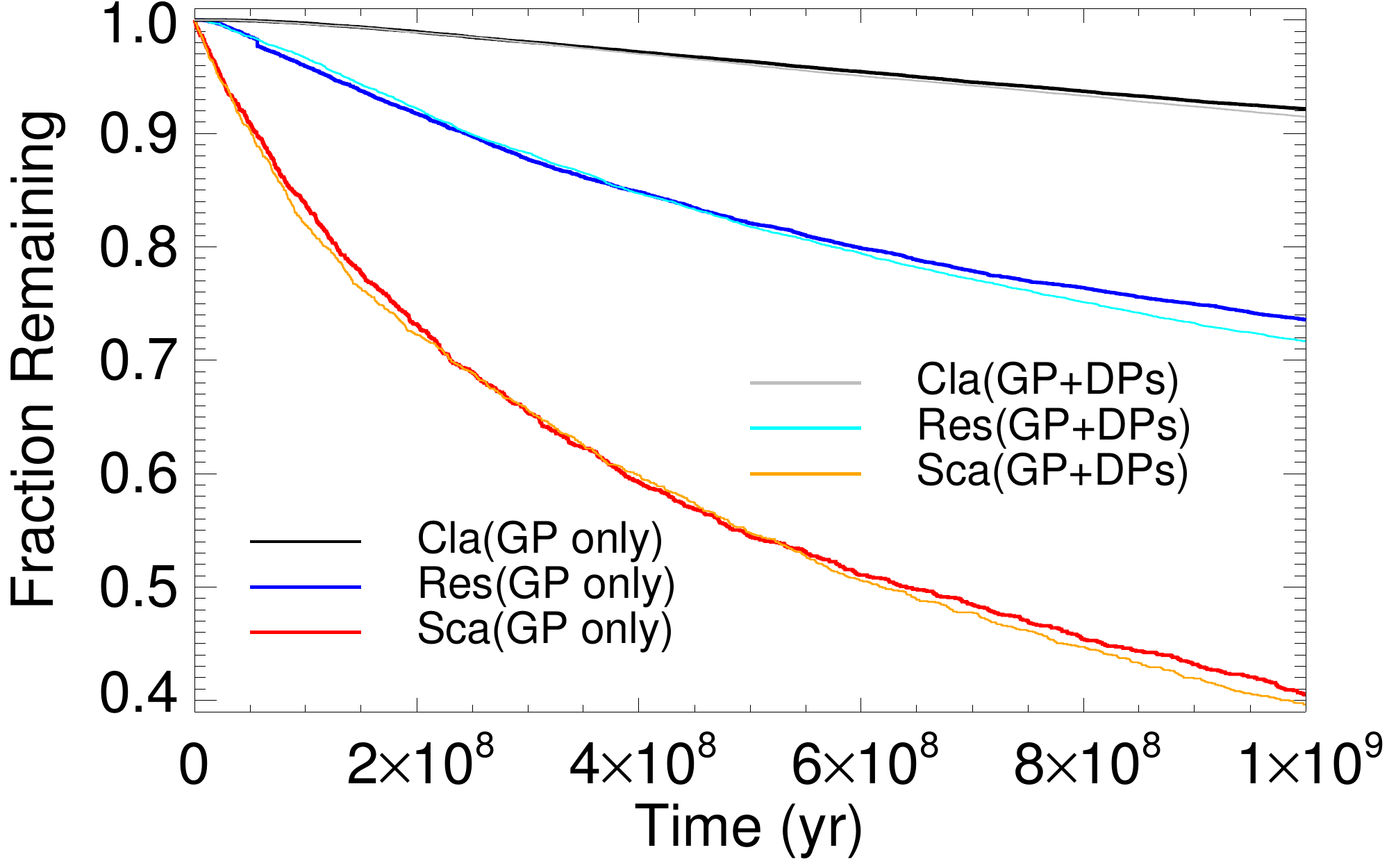}
\caption{Fraction of particles remaining in the Classical, Resonant, and Scattering populations, with and without DPs, as a function of time. The color code is the same as Figure \ref{fig:rmsall}. \label{fig:fracrem}}
\end{figure}

We can further confirm that the increment in the observed number of JFCs in our simulations is a direct product of the dynamical perturbations produced by DPs. To do so we have compared the degree of overall dynamical excitation on the trans-Neptunian cometary particles produced by the giant planets alone, with the excitation produced by the combined effect of the giant planets and the 34 DPs. 

In general, we expect that the inclusion of massive perturbers in the trans-Neptunian region would result in an increase of dynamical excitation of disk particles. From our simulations we observe that such an increase, although small, is significant when compared with the dynamical excitation produced by the giant planets alone. In Fig. \ref{fig:rmsall} we show the root-mean-square (rms) eccentricity and inclination of the L7 model disk particles as a function of time, note that we include all the particles that began the simulation, including the ones that were ejected from the system (and would not be observable in the Solar System).

In the upper panel of Fig. \ref{fig:rmsall}, we can see how the rms eccentricities of the three simulated populations evolve to larger values when DPs are present: 
this is clear for the Scattering and the Resonant populations, although for the Classical population the difference is marginal but noticeable, this is due mainly to the orbital distribution of DPs, given that only 15 DPs are located beyond 50 au, they are unable to exert a more significant perturbation on a large fraction of particles in a vast region of space (44\% of the Classical population have initial $a>50$ au). Note that periods of at least 100 Myr, for the Scattering population, and around 300 Myr for the Resonant population, are required for the effect of DPs to become noticeable. Finally, we can conclude that, as expected, the DPs contribute to globally stir the eccentricities of the objects in the trans-Neptunian region.

In the lower panel of Fig. \ref{fig:rmsall}, we show the evolution of the rms inclinations of the three simulated populations as a function of time. As for the rms eccentricities, the rms inclinations of the Scattering and Resonant populations are larger at the end of the simulations, and the final values are larger with the presence of DPs. For the Classical population a reverse behaviour is observed, as the final value of the rms inclination of the population is slightly smaller than at the beginning of the simulation; this could signify that the L7 model`s inclination is slightly overestimated, since we would expect the Classical population to be nearly in equilibrium; i.e. this could be due to an excess of large inclination particles in the L7 Classical population.

Fig. \ref{fig:rmsrem} is similar to Fig.\ref{fig:rmsall}, but in this case we exclude the particles that are ejected from the simulation (i.e. we only use the particles that would be observed in the Solar system). Similarly as when all particles are counted, we see that the rms eccentricities for the three populations increases as a function of time, with the final values being clearly larger when DPs are present for the Scattering and Resonant populations, and again, only marginally larger for the Classical population.

The rms inclinations of the three populations as a function of time are shown in the lower panel of Fig. \ref{fig:rmsrem}; here we observe the opposite behavior to the one shown by the rms eccentricities (upper panel), with the final rms inclination values being smaller for the Classical and Resonant populations than their initial value at the beginning of the simulations, and where such value is even lower when DPs are present for the Resonant population. In the case of the Scattering population, after an initial increase of the rms inclination lasting up to 300 Myr, the subsequent trend shows a clear decrement of this value, which is more pronounced with the presence of DPs. Although the resonant population shows an overall decrement in the final inclination, there is also a small increase in the first 100-150 Myr. The difference in behavior between Fig. \ref{fig:rmsrem} and Fig.\ref{fig:rmsall} is due to the selection effect caused by the escaping cometary nuclei, i.e. escaping cometary nuclei will have, on average, larger inclinations and eccentricities than the typical disk particle.

Fig. \ref{fig:fracrem} shows the remaining fraction of particles in each population of the L7 model, with and without DPs, as a function of time, using the same color-code as that of the previous figures. It is interesting to note how almost immediately the number of particles of the Scattering population begins to decrease, while at the same time the rms eccentricity and inclination began their increasing trends. Since the scattering particles are the most easily perturbed due to their proximity to Neptune, many of them are lost quickly, either by being ejected from the solar system or by becoming JFCs. 

\section{Source Population Required to Maintain the JFCs in Steady State}
\label{sec:rates}

\begin{deluxetable*}{lcccc}
\tabletypesize{\scriptsize}
\tablecaption{Rates of JFC formation for each of the populations.\label{tbl:rates}}
\tablewidth{0pt}
\tablehead{
\colhead{} & \multicolumn{2}{c}{With DPs} & \multicolumn{2}{c}{Without DPs} \\
\colhead{Population} &  \colhead{Rate (yr$^{-1}$)} & \colhead{Reservoir Size\tablenotemark{a}} & \colhead{Rate (yr$^{-1}$)} & \colhead{Reservoir Size\tablenotemark{a}}}
\startdata
Classical & $3.06\times10^{-11}$ & $(274.3\pm55.5)\times10^6$ & $2.61\times10^{-11}$ & $(321.4\pm65.0)\times10^6$ \\
Resonant & $10.7\times10^{-11}$ & $(78.5\pm15.9)\times10^6$ & $9.54\times10^{-11}$ & $(88.0\pm17.8)\times10^6$ \\
Scattering & $15.0 \times10^{-11}$ & $(55.9\pm11.3)\times10^6$ & $14.6\times10^{-11}$ & $(57.6\pm11.6)\times10^6$
\enddata
\tablenotetext{a}{We include an upper limit to the reservoir population; this limit was obtained by assuming all the new JFCs come from each of the populations, see text.}
\end{deluxetable*}

\begin{figure*}
\plotone{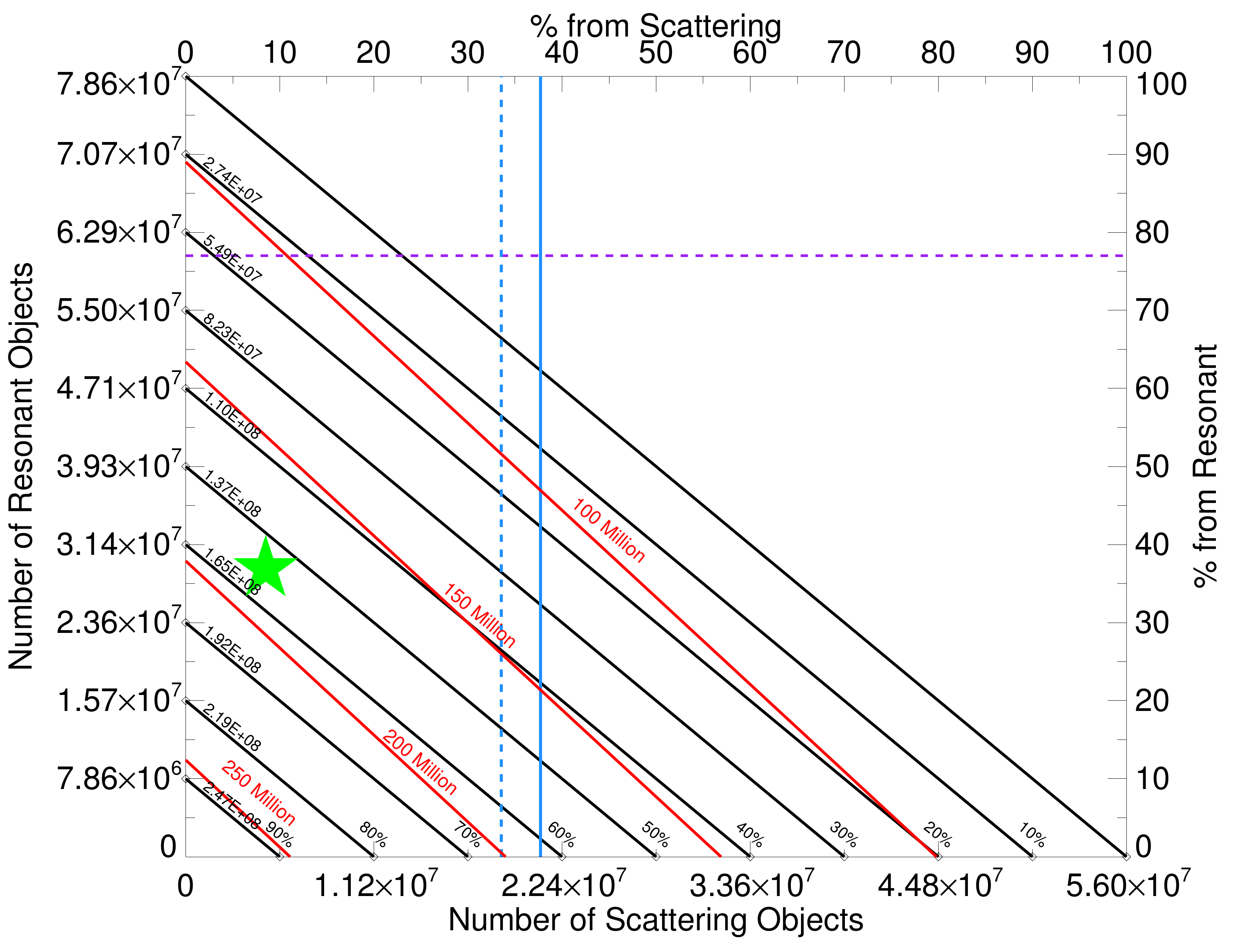}
\caption{Figure representing the different possible combinations of the contributions of each of the three populations to the total JFC production. Each point of the lower left triangle represents one such combination. The fractional contribution of the Scattering and Resonant populations are presented in the upper and right axes respectively, while the remaining fraction, corresponding to the Classical population, is presented in the inner part of the lower axes; the corresponding number of particles required to produce such fractions are presented in the outer lower axis, the outer left axis and the inner left axis for the Scattering, Resonant, and Classical populations, respectively. The red diagonal lines correspond to the sum of the three populations, or the total reservoir of cometary nuclei, the dashed and solid blue lines correspond to the uncorrected and corrected estimation of the scattering population \citep[as estimated by][]{Greenstreet19}, the dashed purple line corresponds to the uncorrected resonant population \citep[as estimated by][]{Greenstreet19}, their corrected Resonant population, as well as their Classical population lie outside the area of the figure, see text. Finally the green star represents our likely distribution as estimated from the L7 particle distribution, see text.\label{fig:popnum}}
\end{figure*}

As mentioned in section \ref{ssec:KB2JFC}, regardless of which populations any particle can be part of (for any period of time), we will label each particle at the beginning of the simulation an keep that label for the entire Gyr.

From our 1 Gyr simulations, we know the efficiency to produce JFCs for each of the three populations in the Kuiper belt, with and without the DPs. We can calculate the injection rate of new visible JFCs, $r_i$, as:
\begin{equation}
r_i = \frac{\njfc}{\tsim\ntot},
\end{equation}
where $\njfc$ is the number of new JFCs produced by each population in our simulations, $\tsim$ is the duration of the simulations, and $\ntot$ is the total number of objects in each simulation. The injection rate efficiencies for each of the populations, are presented in Table~\ref{tbl:rates}, for models both with and without DPs.

By comparing the previous numbers with recent estimations of the required injection rate of new comets, to maintain the population of JFCs in steady state, we can estimate the number of objects in the source populations required to supply such rate. \citet{Rickman17}, by considering the secular (purely dynamical) evolution of JFCs, estimates the injection rate required for a steady state population as $\rr =(8.4\pm1.7)\times10^{-3}$ new comets per year (for cometary nuclei with diameters $D>2$ km). To supply this rate, we have that
\begin{equation}
\rr=\sum \nsrc \rsrc,
\end{equation}
where $\nsrc$ is the number of cometary objects and $\rsrc$ the injection rate for each of the three source populations (with $D>2$ km). We can estimate an upper limit to the number of objects present at each separate population of the Kuiper belt, if we assume that any sub-population contributes with 100\% of the rate given by \citet{Rickman17}. The required reservoir size, for each of the populations, are presented in the third and fifth columns of table \ref{tbl:rates}, for models with and without DPs, respectively.

Currently we do not know, for certain, the exact population count in each of the three sub-populations of the Kuiper belt; at least for the small objects that contribute to the JFCs. In Fig. \ref{fig:popnum}, we present such combinations, where each point in the lower left triangle of the graph represents a specific set of percentages for the contribution from each of the three populations. We consider only the results of the total number of objects estimated when DPs have been included in the simulations, which represents the more realistic model and therefore the more relevant scenario.

The X axis of Fig. \ref{fig:popnum} shows the percentage of JFCs that come from the scattering population (top axis), equivalently it presents the number of objects in the Scattering population required to produce such fraction of the injection rate (lower axis). In a similar way, the Y axis shows the percentage of JFCs originating in the resonant population (right axis), as well as the number of objects required in the Resonant population to obtain such injection rate (left axis). In black diagonal lines, we show the corresponding numbers for the Classical population. Finally, in red diagonal lines, we show the total number of objects of the three populations added together at those locations of the triangle. In this representation, we can set some constrains on the effective contribution of each separate population to the injection rate of JFCs, by combining the number of small objects expected to be present in each of the sub-populations. 

By looking into the L7 model we can estimate the ratio of particles between the Classical, Resonant, and Scattering  populations; which amounts to a 316:61:10 ratio. Given the efficiency of each population this would imply that the percentage of comets arriving from each population to be 
54.6:36.9:8.5 (Fig. \ref{fig:popnum}, green star); 
therefore the number of particles in each population will be 
$149.9\times10^6$, $28.9\times10^6$, and $4.7\times10^6$, 
for the Classical, Resonant, and Scattering populations respectively; for a total population of $1.84\times10^8$ cometary nuclei with $D>2$ km. If we ignore the effect of DPs, the total would be $2.10\times10^8$ cometary nuclei with $D>2$ km; equivalently, the presence of DPs enhances the efficiency of JFC production by about 14.4\% (i.e. 12.6\% of the comets are due to the presence of DPs).

We compared this result with the estimations done by \citet{Greenstreet19}, from the number of craters on the surface of the New Horizons flyby target, 2014 MU$_{69}$; Table 1 of \citet{Greenstreet19} lists an estimate for the number of objects larger than 2 km in diameter in each sub-population. To compare with our estimates of the populations we proceed as follows: for the Scattering disk objects we multiply the number that Greenstreet et al. label as ``Scattering objects with $15\le a\le200$ au'' by a factor of 1.055 since, 12 out of 242 (5\%) of the objects that become visible JFCs, have an initial semimajor axis larger than 200 au, equivalently 146 out of 1612 (9.1\%) of the total set of Scattering disk particles have an initial semimajor axis larger than 200 au. For the Resonant population Greenstreet et al. only provide estimations for the 3:2, 2:1, 5:2, and 7:4 MMRs; we note that, in our simulations, these resonances only account for approximately 68.5\% of the number of particles on all the MMRs and correct accordingly. Finally, for the Classical population, we combine Greenstreet's ``Classical Inner'', ``Classical Main H'', ``Classical Main S", ``Classical Main K'', and ``Classical Outer'' (``H'', ``S'', and ``K'' stand for hot, stirred, and kernel respectively) into one single number to compare with our Classical population.

The total set of objects predicted by \citet{Greenstreet19} would imply a JFC injection rate 2.9 times greater than the one required by our steady state assumption: the scattering population provides $\sim38\%$ of the required injection rate (blue line in Fig. \ref{fig:popnum}), the resonant population provides $\sim110\%$ of the required injection rate (out of boundaries), and the classical population provides $\sim145\%$ of the required injection rate (also out of boundaries). 

Within our model, the rate of comets produced requires a well-defined number of objects in the reservoirs, in order to supply the renewal rate of \citet{Rickman17}. We find that the requirements for the reservoir size populations are within sensible limits with recent observational estimations, i.e. the discrepancy between theory and observational-based predictions is only of a factor $\sim$ 3, and not of orders of magnitude. Given the uncertainties involved in both the observations and theoretical assumptions, this factor 3 seems not too worrying and even more, the current numerical integrations are able to provide more than enough new objects for the population of JFCs to remain in steady state. This means that no additional factors (such as still undiscovered planets) are necessary to account for the origin of the low-inclination comets of the solar system.

\section{Conclusions}
\label{sec:conc}

In this work we have explored the long-term dynamical evolution, during 1 Gyr, of an un-biased orbital representation of cometary nuclei in the trans-Neptunian region, which includes the Classical Kuiper belt, a Resonant population distributed among 11 MMRs, and the Scattering disk: the so-called L7 model. We analyzed the evolution of the L7 model under the influence of the giant planets alone, from Jupiter to Neptune, and under the effect of the giant planets together with the 34 largest TNOs observed to date, those with and absolute visual magnitude $H_V<4$.

We measured the number of Jupiter Family Comets that are produced in the numerical simulations; where we measured the effect, that the presence of the 34 largest TNOs in the Kuiper belt, has in the enhancement of the production of JFCs. The overall increment in the number of JFCs produced by the presence of DPs accounts for up to 12.6\%, when compared with the number produced by the giant planets alone. This enhancement shows that, although modest, the influence of DP-sized objects in the outer Solar System is a non-negligible factor on the secular evolution of the Kuiper belt; therefore, the gravitational influence of the objects in the bright end of the size distribution, should be taken into account when considering the long-term evolution of the Solar System debris belts, as well as when considering the evolution of extrasolar debris disks in general. 

Regarding the effect of DPs over each the three populations of the Kuiper belt separately, we found that: for the Scattering population, DPs have only a small effect in increasing the number of Crossers, NIPs, and JFCs; for the Classical population the enhancement produced by DPs is quite large, accounting for up to 17.2\% more JFCs than the ones produced by the giant planets alone; while for the Resonant population, the increment in the number of JFCs is approximately 12\%, but depends strongly on the specific resonance.

When studying each resonance separately in the presence of DPs, we found that the Plutinos alone contribute about 45\% of all the originally resonant JFCs. Surprisingly, the second most important MMR that contributes to the JFCs is the 5:3 ($\sim15\%$), despite being only the fourth most populated resonance and having less than half the number of objects than the 5:2 MMR. In this regard, the lack of new comets coming from the 5:2 MMR, despite the large population that it is estimated to harbor, can be understood in terms of the overall stability of this resonance, recently demonstrated by \citet{Malhotra18}. 

Given the injection rates of new visible JFCs produced by our simulations, we have estimated the number of objects required to maintain the observed population of JFCs; this has let us provide upper limits to the number of objects, with diameters larger than 2 km, in each population of the Kuiper belt. These limits are broadly in accordance with recent independent estimations; they are also in line with current trends, which predict a lower number of small-sized objects in the trans-Neptunian region than previously though. In this sense, as few as 184 million objects ---in the whole Kuiper belt--- are able to supply the inner Solar System with the new comets required to maintain the observed steady state population of JFCs.

\acknowledgments

We acknowledge the referee, Luke Dones, for a helpful review. We thank K. Volk, W. Fraser, and R. Pike for helpful comments and discussions. BP, AP, and MAM acknowledge grants DGAPA-PAPIIT-IN101918 and CONACyT Ciencia B\'asica grant 255167. AP acknowledges grant DGAPA-PAPIIT IG-100319. We want to dedicate this work to the loving memory of Barbara Pichardo, who started very enthusiastically this collaboration but unfortunately is no longer with us.

\appendix

\section{The 34 Largest TNOs}
\label{apdx:dps}

Current observational works can provide a trustworthy estimation of the size of a bright object, but for many of them, the mass and density are far less well constrained quantities, if at all. For the cases of TNOs with observed satellites, the mass can be estimated by using Kepler's third law. The project ``TNOs are cool" has provided data for the size, absolute magnitude, and geometric albedo of a large sample of TNOs; they do this by making use of the infrared and optical data points in the spectral energy distributions of the TNOs, obtained with the {\it Hershel} and {\it Spitzer} space telescopes, and fitting a thermal model to calculate the disk-integrated thermal emission to the observations \citep{Stansberry08,Muller10,Santos12}. Other works rely on precise orbital determinations for some TNOs, in order to predict stellar occultations from which to extract fine measurements of the size and mass of the objects. For Pluto and its satellite system, the New Horizons space mission has provided the best data available.

In Table \ref{tbl:knownDPs} we list the data and references, for the 13 objects for which a trustworthy measure of the size and mass (and therefore density) are available in the literature. In Table \ref{tbl:unknownDPs} we present the data for another 14 objects for which only the size can be provided by the observations, but neither density nor mass are available. Finally, Table \ref{tbl:albDPs} presents the data for 7 objects for which only a measure of the absolute magnitude is provided by the MPC.

To assign a size to the 7 objects with only known $H_V$, we used the absolute magnitudes, as listed in the MPC, of the remaining 27 objects and plotted those values against their geometric albedos, $p$, as given in several references, mainly from the ``TNOs are cool'' series of papers. We made a fit to the 27 objects with published values of $p$, given by:
\begin{equation}
p=
\begin{cases}
0.0403H_V^2-0.2588H_V+0.5558 & \text{if } H_V \le 3.21 \\
0.1403 & \text{if } H_V > 3.21
\end{cases},
\label{eq:pvsh}
\end{equation}
then we assign a random value of $p$ to the 7 objects with no published data, within one $\sigma$ of the fitted curve. Once we have set the values for $p$ and $H_V$, we use the relation between size, absolute magnitude, and geometric albedo \citep{Harris97} given by
\begin{equation}
R=664.5 {\it km}\, {p}^{-0.5}10^{-H_V/5},
\label{eq:dph}
\end{equation}
to finally assign a radius, $R$, to the 7 objects.

Similarly, in order to assign a density to the 21 objects without known data, we have used the measured densities of the 13 objects with known density (those in Table \ref{tbl:knownDPs}). Again, we made a fit to this distribution of densities versus sizes for these 13 objects, which is given by:
\begin{equation}
\rho=\left[\left(\frac{R}{220 {\it km}}\right)^{-3}+(2.1)^{-3}\right]^{-1/3},
\label{eq:rvsr}
\end{equation}
where density, $\rho$, is given in gr cm$^{-3}$. In this manner, we are able to randomly assign a corresponding density for the remaining 21 objects, according to their radius, within one $\sigma$ of the fitted curve.

\begin{figure}
\plotone{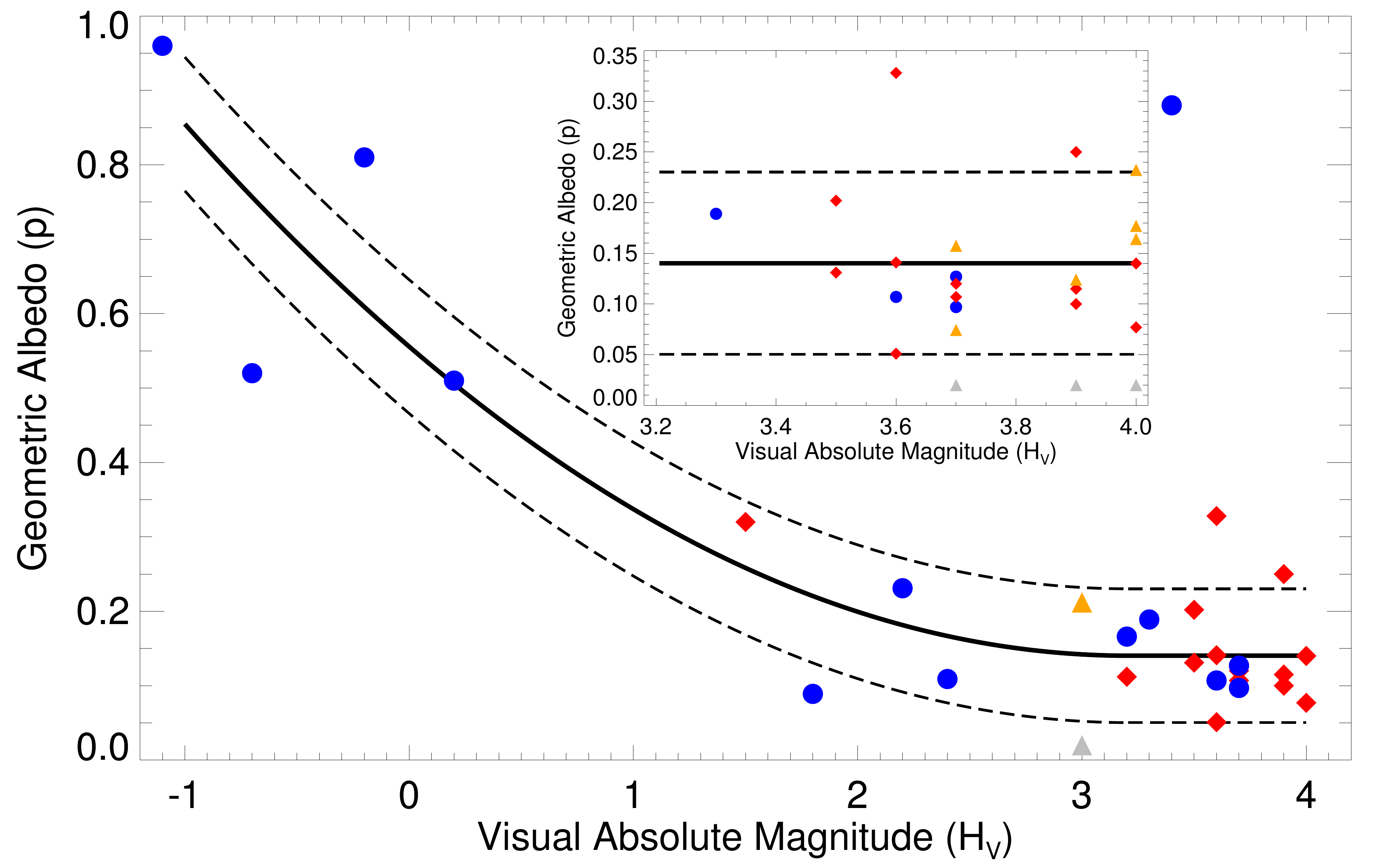}
\caption{Geometric albedo vs. absolute magnitude for all objects in Tables \ref{tbl:knownDPs}, \ref{tbl:unknownDPs}, and \ref{tbl:albDPs}. Blue circles indicate the 13 objects for which all data is known (those in Table \ref{tbl:knownDPs}), red diamonds indicate the 14 objects with known radius and geometric albedo (those in Table \ref{tbl:unknownDPs}). Gray triangles show the absolute magnitude of the 7 objects in Table \ref{tbl:albDPs}, while the orange triangles indicate the random values of the albedos for the latter, obtained by using the fitted trend to the blue circles plus red diamonds, given by equation \ref{eq:pvsh}. The inlay is a zoomed region beyond $H_V=3.2$ for clarity. The one $\sigma$ region around the fitted trend is delimited by the gray shaded area.\label{fig:DPsHvsp}}
\end{figure}

Fig. \ref{fig:DPsHvsp} shows the distribution of $p$ {\it vs.} $H_V$, for the objects with these two data known from observations, and a fit is made to the distribution in order to assign a random albedo to the remaining 7 objects without data. The inlay in the Figure is a zoom to better visualize the region beyond $H_V = 3.21$.

\begin{figure}
\plotone{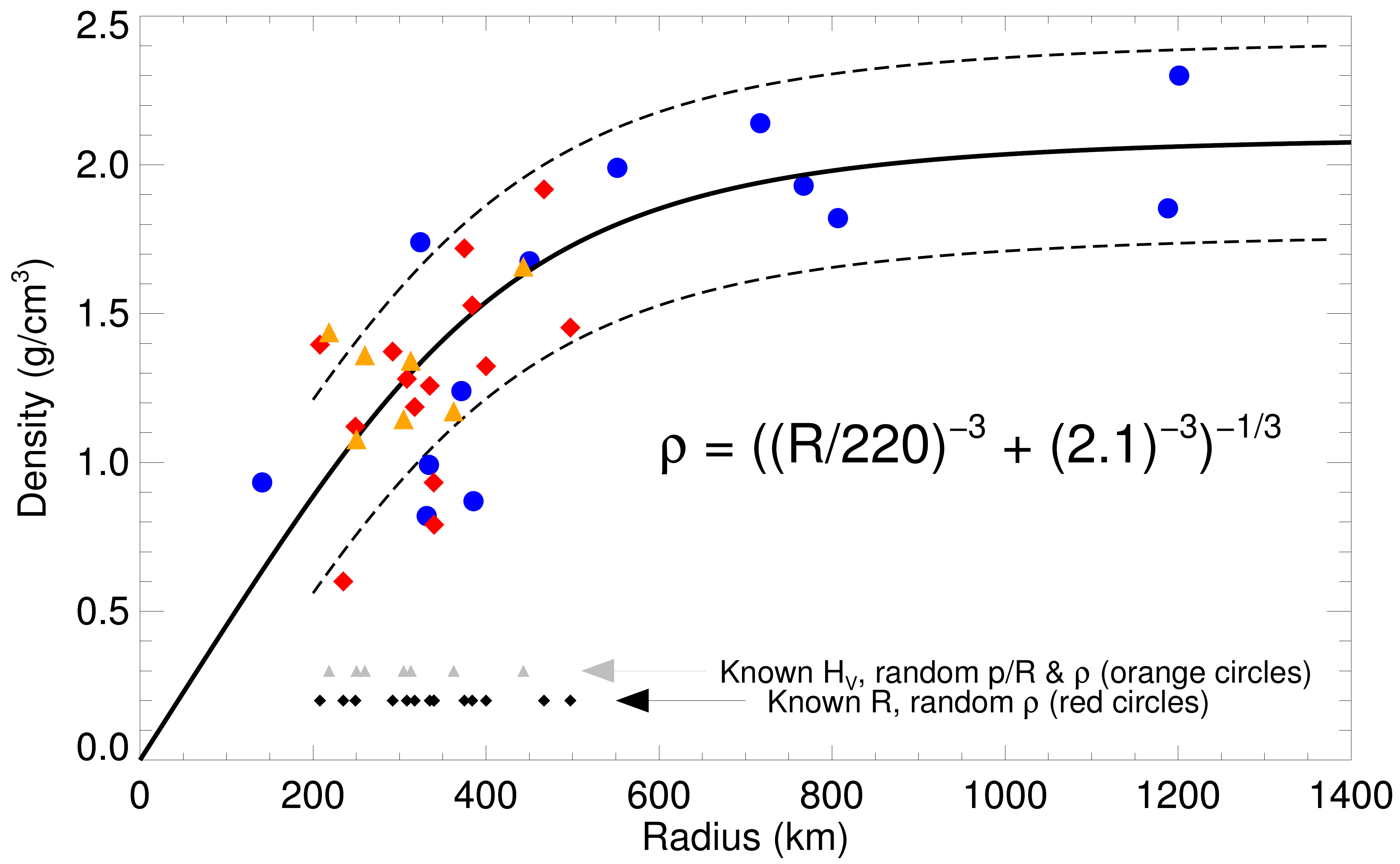}
\caption{Density vs. radius for all objects in Tables \ref{tbl:knownDPs}, \ref{tbl:unknownDPs}, and \ref{tbl:albDPs}. Blue circles are for objects with measured densities (from Table \ref{tbl:knownDPs}). Smaller black diamonds and gray triangles show the size of objects in Tables \ref{tbl:unknownDPs} and \ref{tbl:albDPs}, respectively, for which a random density is assigned by using the fitted trend given by equation \ref{eq:rvsr}. Red diamonds are for objects in Table \ref{tbl:unknownDPs}, while orange triangles are for objects in Table \ref{tbl:albDPs}. The gray shaded area delimits the one $\sigma$ around the fitted curve.\label{fig:DPsrvsd}}
\end{figure}

Finally, Fig. \ref{fig:DPsrvsd} shows the distribution of densities versus sizes, where a fitting trend to the 13 objects will full known data is performed in order to randomly assign a corresponding density for the remaining 21 objects, according to their radius, within one $\sigma$ of the fitted curve.

\begin{deluxetable}{lccc}
\tabletypesize{\scriptsize}
\tablecaption{Large trans-Neptunian objects with known data for radius, mass, and/or density.\label{tbl:knownDPs}}
\tablewidth{0pt}
\tablehead{
\colhead{Name} & \colhead{$R$ (km)} & \colhead{$\rho$ (gr cm$^{-3}$)} & \colhead{$M$ ($\times 10^{-3}$ M$_\oplus$)}
}
\startdata
Eris\tablenotemark{1} & 1200.0 & 2.3 & 2.7956 \\
Pluto\tablenotemark{2} & 1188.3 & 1.854 & 2.4467 \\
Makemake\tablenotemark{3} & 717.0 & 2.14 & 0.5531 \\
Haumea\tablenotemark{4} & 806.8 & 1.821 & 0.6706 \\
2007 OR$_{10}$\tablenotemark{5} & 767.2 & 1.9273 & 0.6110 \\
Orcus\tablenotemark{6} & 450.0 & 1.676 & 0.1073 \\
Quaoar\tablenotemark{7} & 551.7 & 1.99 & 0.2343 \\
Varda\tablenotemark{8} & 371.5 & 1.24 & 0.0446 \\
2007 UK$_{126}$\tablenotemark{9} & 324.0 & 1.74 & 0.0415 \\
2002 TX$_{300}$\tablenotemark{10} & 143.0 & 0.933 & 0.0018 \\
2002 UX$_{25}$\tablenotemark{11} & 332.0 & 0.82 & 0.0209 \\
Varuna\tablenotemark{12} & 334.0 & 0.992 & 0.0259 \\
2003 AZ$_{84}$\tablenotemark{13} & 385.5 & 0.87 & 0.0349 \\
\enddata

\tablecomments{We are interested in mass, $M$, and density, $\rho$, as input data for the simulations. However, for most objects, the radius, $R$, is the better constrained quantity. Here we quote only the radius we used in order to calculate the mass or the density for each object (i.e. without any associated uncertainty; such uncertainties can be found in the references given for each object). In some cases, the density publicly available is only an upper or lower limit, according to the available data. In those cases we use the quoted value as the density for the object. See the notes and references for details about each object. In all cases we consider spherical shapes, in order to calculate a radius, density, or mass, depending on the available data.}

\tablenotetext{1}{All $R$, $\rho$, and $M$ are measured \citep{Brown08}.}
\tablenotetext{2}{All $R$, $\rho$, and $M$ are measured with high accuracy \citep{Stern18}. For the simulations we used the added mass of Pluto + Charon, and their center of mass as the origin of a unique object with an average density value of 1.781 gr cm$^{-3}$.}
\tablenotetext{3}{Density for Makemake is not well constrained. We use the value 2.14, which is only a model-dependent lower limit \citep{Brown13}, in order to estimate a mass of a spherical object with the measured radius.}
\tablenotetext{4}{Density is estimated to range from [1.757,1.885] gr cm$^{-3}$ \citep{Ortiz17}; we used the average value and the mass given by \citet{Ragozzine09} to estimate the radius. Note, however, that Haumea has a strong triaxial shape.}
\tablenotetext{5}{The mass and size of the object are estimated in \citet{Kiss17}. We used the central value in the range and assumed a spherical body in order to derive a density.}
\tablenotetext{6}{Quoted are the middle values for both $R$ and $\rho$, among the possible ranges given in \citet{Barr16}. Mass is estimated here from such middle values assuming a spherical body.}
\tablenotetext{7}{Radius is estimated here considering a spherical object with mass given by \citet{Fraser13} and density as quoted by \citet{Barr16}.}
\tablenotetext{8}{Mass and density are measured and given by \citep{Grundy15}. Here we estimated the radius for a spherical object.}
\tablenotetext{9}{The radius and an upper limit density are estimated in \citet{Benedetti16} from a stellar occultation. Mass is derived here assuming a spherical object.}
\tablenotetext{10}{The radius is measured and water ice density is assumed in \citet{Elliot10}. Mass is thus derived here. Note that this object belongs to the Haumea collisional family \citep[see for example][]{Lykawka12}.}
\tablenotetext{11}{We use the radius given by \citet{Brown13b}, who considers a smaller value than the one measured by \citet{Fornasier13}, as adequate for the primary object alone. Density was measured by \citet{Brown13b}. We derive the mass assuming a spherical body.}
\tablenotetext{12}{Density is given by \citet{Lacerda07} and radius is estimated by \citet{Lellouch13}. Mass is estimated from these values.}
\tablenotetext{13}{Density is derived from the rotational light curve in \citet{DiasO17}. We use the radius given in \citet{Mommert12}, considering a spherical body, to derive the mass.}

\end{deluxetable}

\begin{deluxetable}{lcccl}
\tabletypesize{\scriptsize}
\tablecaption{Large trans-Npetunian objects with only known data for the radius.\label{tbl:unknownDPs}}
\tablewidth{0pt}
\tablehead{
\colhead{Name} & \colhead{$R$ (km)} & \colhead{$\rho$ (gr cm$^{-3}$)} & \colhead{$M$ ($\times10^{-3}$ M$_\oplus$)} & \colhead{Ref} 
}
\startdata
Sedna & 497.5 & 1.4532 & 0.1254 & 1 \\
2002 AW$_{197}$ & 384.0 & 1.5277 & 0.0606 & 2 \\
2014 UZ$_{224}$ & 317.5 & 1.1865 & 0.0266 & 3 \\
2005 UQ$_{513}$ & 249.0 & 1.1206 & 0.0121 & 2 \\
Ixion & 308.5 & 1.2811 & 0.0263 & 4 \\
2002 MS$_{4}$ & 467.0 & 1.9176 & 0.1369 & 5 \\
2005 QU$_{182}$ & 208.0 & 1.3957 & 0.0088 & 6 \\
2015 RR$_{245}$ & 335.0 & 1.2577 & 0.0331 & 7 \\
2005 RN$_{43}$ & 339.5 & 0.9326 & 0.0255 & 5 \\
2002 TC$_{302}$ & 292.0 & 1.3725 & 0.0239 & 8 \\
2015 KH$_{162}$ & 400.0 & 1.3236 & 0.0594 & 9 \\
2010 EK$_{139}$ & 235.0 & 0.6002 & 0.0054 & 1 \\
2004 GV$_9$ & 340.0 & 0.7910 & 0.0218 & 5 \\
2010 JO$_{179}$ & 375.0 & 1.7194 & 0.0635 & 10 \\
\enddata

\tablecomments{For the objects in this table only the radius is known from observations. We assigned a random density by fitting a trend to the 13 objects with known density, then we derive the mass of the object assuming it is spherical. See also Fig. \ref{fig:DPsrvsd}}

\tablerefs{
(1) \citet{Pal12}; (2) \citet{Vilenius14}; (3) \citet{Gerdes17}; (4) \citet{Lellouch13}; (5) \citet{Vilenius12}; (6) \citet{Santos12}; (7) \citet{Bannister16}; (8) \citet{Fornasier13}; (9) \citet{Sheppard16}; (10) \citet{Holman18}.
}

\end{deluxetable}

\begin{deluxetable}{lccccc}
\tabletypesize{\scriptsize}
\tablecaption{Large trans-Neptunian objects with only visual absolute magnitude, $H_V$, known.\label{tbl:albDPs}}
\tablewidth{0pt}
\tablehead{
&  &  & \colhead{$R$} & \colhead{$\rho$} & \colhead{$M$}\\
Object & \colhead{$H_V$} & \colhead{$p$} & \colhead{(km)} & \colhead{(gr cm$^{-3}$)} & \colhead{($\times10^{-3}$ M$_\oplus$)} 
}
\startdata
2013 FY$_{27}$ & 3.0 & 0.21 & 362.5 & 1.1722 & 0.0391 \\
2014 EZ$_{51}$ & 3.7 & 0.07 & 443.0 & 1.6585 & 0.1012 \\
2010 RF$_{43}$ & 3.7 & 0.15 & 305.0 & 1.1458 & 0.0227 \\
2003 OP$_{32}$ & 3.9 & 0.12 & 313.0 & 1.3422 & 0.0288 \\
2012 VP$_{113}$ & 4.0 & 0.17 & 250.0 & 1.0785 & 0.0118 \\
2010 KZ$_{39}$ & 4.0 & 0.16 & 260.0 & 1.3606 & 0.0167 \\
2014 WK$_{509}$ & 4.0 & 0.23 & 218.5 & 1.4378 & 0.0105 \\
\enddata

\tablecomments{For the objects in this table only an absolute magnitude, $H_V$, is listed on the MPC website. We derive a geometric albedo, $p$, by fitting a trend to the 27 objects in Tables \ref{tbl:knownDPs} and \ref{tbl:unknownDPs} for which data for $p$ is available in the literature; we then randomly assign an albedo within one $\sigma$ of the trend. A radius is derived by using equation \ref{eq:dph}, and finally a random density by using the fitted trend shown in Fig. \ref{fig:DPsrvsd}. For the masses we assumed spherical objects in all cases.}

\end{deluxetable}

\bibliographystyle{aasjournal}
\bibliography{comets}

\clearpage


\end{document}